\documentclass[sigconf,authorversion,nonacm, table]{acmart}
\AtBeginDocument{%
  \providecommand\BibTeX{{%
    \normalfont B\kern-0.5em{\scshape i\kern-0.25em b}\kern-0.8em\TeX}}}

\usepackage{filecontents,lipsum}
\usepackage{amsmath, nccmath}
\usepackage{algorithmic}
\usepackage{array}
\usepackage{bm}
\usepackage{graphicx}
\usepackage{makecell}
\usepackage{multirow}
\usepackage{subcaption}
\usepackage{tabularx,booktabs}
\usepackage{textcomp}
\usepackage{xcolor}
\DeclareMathOperator*{\argmax}{\arg\!\max}
\DeclareMathAlphabet{\mathbb}{U}{bbold}{m}{n}
\DeclareUnicodeCharacter{2212}{-}

\makeatletter
\newcommand{\thickhline}{%
    \noalign {\ifnum 0=`}\fi \hrule height 1pt
    \futurelet \reserved@a \@xhline
}
\newcolumntype{"}{@{\hskip\tabcolsep\vrule width 1pt\hskip\tabcolsep}}

\makeatother

\newcommand{\newedit}[1]{{#1}}
\begin{document}

\setlength{\abovedisplayskip}{3pt}
\setlength{\belowdisplayskip}{3pt}

\title{Within-basket Recommendation via \\
Neural Pattern Associator}

\settopmatter{authorsperrow=4}

\author{Kai Luo $^{\dagger}$}
\affiliation{%
  \institution{Loblaw Digital}
  \city{Toronto}
  \state{ON}
  \country{Canada}
  }
\email{kai.luo@loblaw.ca}

\author{Tianshu Shen $^{\dagger}$}
\affiliation{%
  \institution{Loblaw Digital}
  \city{Toronto}
  \state{ON}
  \country{Canada}
  }
\email{tina.shen@loblaw.ca}

\author{Lan Yao}
\affiliation{%
  \institution{Loblaw Digital}
  \city{Toronto}
  \state{ON}
  \country{Canada}
  }
\email{lan.yao@loblaw.ca}

\author{Ga Wu}
\affiliation{%
  \institution{Dalhousie University}
  \city{Halifax}
  \state{NS}
  \country{Canada}
  }
\email{ga.wu@dal.ca}

\author{Aaron Liblong $^{\ddagger}$}
\affiliation{%
  \institution{Loblaw Digital}
  \city{Toronto}
  \state{ON}
  \country{Canada}
  }
\email{aaron.liblong@loblaw.ca}

\author{Istvan Fehervari $^{\ddagger}$}
\affiliation{%
  \institution{Loblaw Digital}
  \city{Toronto}
  \state{ON}
  \country{Canada}
  }
\email{istvan.fehervari@loblaw.ca}

\author{Ruijian An}
\affiliation{%
  \institution{Loblaw Digital}
  \city{Toronto}
  \state{ON}
  \country{Canada}
  }
\email{ruijian.an@loblaw.ca}

\author{Jawad Ahmed $^{\ddagger}$}
\affiliation{%
  \institution{Loblaw Digital}
  \city{Toronto}
  \state{ON}
  \country{Canada}
  }
\email{jawad.ahmed01@loblaw.ca}

\author{Harshit Mishra}
\affiliation{%
  \institution{Loblaw Digital}
  \city{Toronto}
  \state{ON}
  \country{Canada}
  }
\email{harshit.mishra01@loblaw.ca}

\author{Charu Pujari}
\affiliation{%
  \institution{Loblaw Digital}
  \city{Toronto}
  \state{ON}
  \country{Canada}
  }
\email{charu.pujari@loblaw.ca}


\begin{abstract}
\emph{Within-basket recommendation}~(WBR) refers to the task of recommending items to the end of completing a non-empty shopping basket during a shopping session. 
% This task presents a research challenge in jointly modeling three factors, namely 
% 1) co-existence of multiple shopping intentions, 
% 2) multi-granularity of such intentions, and 
% 3) interleaving behavior (switching intentions) in a shopping session.
% weak order-dependence of known items in baskets. \tina{sounds like solution, could replace with user intentions interleaving.} 
While the latest innovations in this space demonstrate remarkable performance improvement %over classic association-rule-based methods 
on benchmark datasets, they often overlook the complexity of user behaviors in practice, such as
1) co-existence of multiple shopping intentions, 
2) multi-granularity of such intentions, and 
3) interleaving behavior (switching intentions) in a shopping session.
%\textcolor{red}{seldom model one-to-multiple}
%of these factors, % (e.g., by falling back to exploit the pairwise similarity of products), 
%resulting in suboptimal performance in landing WBR services. 
This paper presents Neural Pattern Associator~(NPA), a deep item-association-mining model that explicitly models the aforementioned factors. Specifically, inspired by vector quantization, the NPA model learns to encode common user intentions (or item-combination patterns) as quantized representations (a.k.a. codebook), which permits identification of users' shopping intentions via attention-driven lookup during the reasoning phase. This yields coherent and self-interpretable recommendations. 
% We have evaluated the NPA model's performance at the within-basket recommendation in several large datasets, in the domains of grocery e-commerce (shopping basket completion) and music (playlist extension). 
We evaluated the proposed NPA model across multiple extensive datasets, encompassing the domains of grocery e-commerce (shopping basket completion) and music (playlist extension), where our quantitative evaluations show that the NPA model significantly outperforms a wide range of existing WBR solutions, reflecting the benefit of explicitly modeling complex user intentions.
% \tina{should we put this in footnote instead?} \kai{we can put the url anywhere as long as the reviewers are not blind :). Apparently, the way we put url in the ICDM submission is not obvious enough.}

\def\thefootnote{$\dagger$}\footnotetext{These authors contributed equally to this work}

\def\thefootnote{$\ddagger$}\footnotetext{Contributions were made while the author was at Loblaw Digit.}

\end{abstract}





\maketitle

\section{Introduction}
% Recommender systems one of the essential services of e-commerce (e.g., Instacart, Amazon, etc.) and entertainment (e.g., Spotify, Netflix, etc.) which ease clients' effort for finding items they might be interested in from wide variety of contents/products in service provider' inventory. 
% In many of these scenarios, such as grocery shopping, users select multiple items in a single transaction, which corresponds to a shopping basket. 

Within-basket recommendation~(WBR)~\cite{saxena2015frequent,wan2018representing, ariannezhad2023personalized} is a specific type of recommendation task that aims to ``complete'' a shopping basket with a set of relevant items, where the combination of already-present and recommended items should form a cohesive set that aligns with the user's intentions. 
In light of the objective of maintaining basket coherence, 
% within-basket recommendation 
WBR can be viewed as a set-expansion problem, where intra-set item association is the core modeling factor. Early solutions, such as Apriori~\cite{agrawal1994fast}, confronted the WBR task by mining and applying item association rules. However, with the growth of variety in e-commerce inventories, these classic methods diminished in practicality due to scalability issues.

% Early solutions such as Apriori~\cite{agrawal1994fast}, \newedit
% {initially tackled the WBR task by mining and applying item association rules but scalability issues arose with the growth of variety in e-commerce inventories.}

% these classic methods diminished in practicality due to scalability issues. 
% Consequently, machine learning (ML) 
% techniques began to dominate the space. 
% To date, ML approaches have typically assumed that items within a basket are semantically similar~\cite{le2017basket, wan2018representing, liu2020basconv, xu2020knowledge}. 
% Such solutions can be broadly categorized based on model architecture into two, 
% In addition, based on their architecture or framework choices, the existing ML-based solutions can be broadly categorized into two tracks, 
% namely matrix decomposition~\cite{grbovic2015commerce, barkan2016item2vec, wan2018representing, xu2020knowledge} 
% and graph convolution~\cite{liu2020basket, liu2020basconv}. 
% Matrix-decomposition-based models
% need a key sentence here
Machine-learning (ML) solutions greatly improved the scalability of WBR systems by parameterizing item compatibility.
Representation-learning-based approaches~\cite{grbovic2015commerce, barkan2016item2vec, xu2020knowledge}, 
% wan2018representing,
as an example, employ self-supervised training schema to estimate item and basket (or user) representations that implicitly encode the compatibility of the two. 
%\tina{the citations in the previous sentence are all self-supervised?}  
% Despite their simplicity, matrix-decomposition-based approaches are robust in their handling of sparse observations in data. 
Another example is graph-convolution-based models~\cite{liu2020basket, liu2020basconv}, which mine complex associations among items and distribute mined associations into item representations~\cite{liu2020basconv}. Unfortunately, many of the aforementioned solutions were primarily invented to advance academic discovery, relying on numerous laboratory assumptions (e.g., assuming that all incomplete baskets are observed during the training phase with unique shopping intentions~\cite{liu2020basconv, liu2020basket}, as pointed out in~\cite{ariannezhad2023personalized}), thereby limiting their usefulness and practical applicability. 
% \textcolor{red}{practice.}
% real production environments. 
% The latest graph-based solutions have explored modeling multiple combination patterns~\cite{liu2020basket} or their corresponding granularities~\cite{guo2022learning}, but such research is in an early stage, where 
% However, strong laboratory assumptions have been made (e.g., assuming all incomplete baskets are observed during the training phase~\cite{liu2020basconv, liu2020basket}, as pointed out in~\cite{ariannezhad2023personalized}). 
% Such models are further limited in their applicability to WBR; their static nature is ill-suited to modeling the dynamic evolution of a basket during a shopping session.

\begin{figure*}[tbh]
    \centering
        \includegraphics[width=0.9\textwidth]{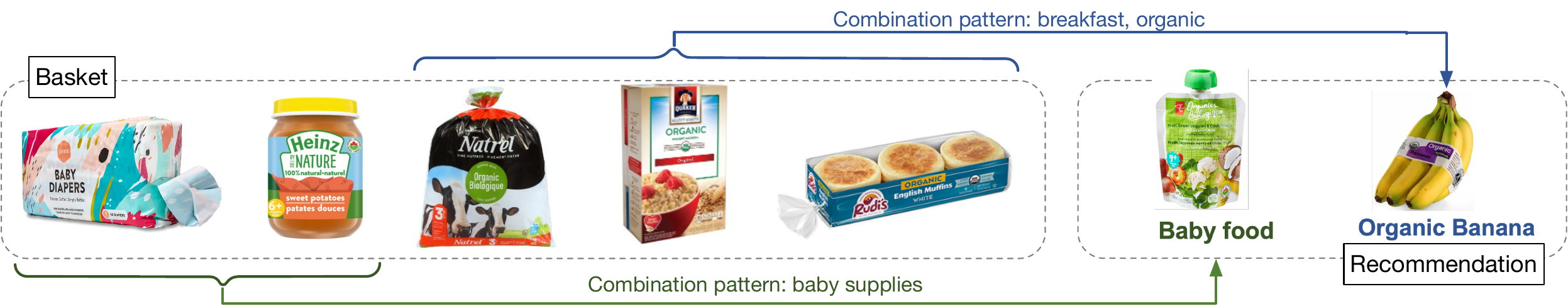}
    \caption{A high-level illustrative figure on how Neural Pattern Associator (NPA) makes its recommendations. Given baby diapers, baby food jar, organic milk, organic instant oatmeal, and organic English muffin, NPA identifies two combination patterns (baby supplies and organic breakfast). The product add-to-cart sequence has a minor impact on the NPA model as it supports intention interleaving naturally, as the effect of codebook lookup. As a result, NPA recommends fruity baby food and organic bananas as complementary to the products in the basket.
    %\tina{I think we need to add text label for the input items as well, as the item fonts are pretty small now.}
    % {\color{red} This is ugly ..}
    % \tina{Example figure, to be replaced}
    }
\label{fig:example}
\vspace{-5mm}
\end{figure*}

Recent work has shifted its focus toward more practical settings, specifically by incorporating user intentions into the modeling process. In particular, items within a shopping basket could reflect multiple shared user intentions (referred as \textit{combination patterns} in this work)~\cite{liu2020basket, shen2023temporal, wang2021learning, chen2022intent}. For a shopping basket (Figure~\ref{fig:example}), a subset of items, \{\emph{baby diapers}, \emph{baby food}\}, indicates the user's intention of buying baby supplies, whereas the remaining items \{\emph{organic milk}, \emph{organic oat meal}, \emph{organic English muffin}\} reflects the user's intention to prepare a healthy breakfast. 
% no need to talk about attention here -- Wuga
% Among these works, the attention mechanism~\cite{vaswani2017attention} is typically used for multi-intent extraction.
% \tina{could extend to describe existing works, and relate with the attention mechanism, but are in slightly different settings.}
However, existing works often overlook the complexity of user behaviors in real shopping events. In fact, a higher-level shopping intention (or a complex item combination pattern) could also be satisfied with respect to multiple granularity levels and focused aspects~\cite{guo2022learning}. For instance, a shopping cart before Christmas Eve includes ingredients for dinner, toys or gifts for children, and even Christmas decorations, 
each representing a distinct lower-level shopping intention.
% each of which could be a lower level shopping intention. 
% For example, one could interpret a basket containing \{\emph{spaghetti}, \emph{spicy salami}, \emph{pesto}, \emph{parmesan}, \emph{olives}, \emph{tiramisu}\} along multiple aspects, such as flavor (e.g., spicy, salty, sweet); meal type (e.g., main courses, appetizers, desserts); and culture (e.g., Italian). 
In addition, the noisy order of adding items to a basket can introduce an additional layer of modeling challenge as user intentions could be interleaving~\cite{han2021multiple, hendriksen2020analyzing}.

In this paper, we present Neural Pattern Associator (NPA) --- a general inductive framework particularly designed to address the aforementioned intention modeling challenges of WBR.
% within-basket recommendation. 
% In this paper, we present Neural Pattern Associator (NPA), a general inductive framework to solve WBR challenges, that combines the Attention model~\cite{vaswani2017attention}, Vector Quantized Representation~\cite{agustsson2017soft, theis2017lossy, van2017neural}, and Any-order Autoregressive Training~\cite{uria2014deep, strauss2021arbitrary, hoogeboom2021autoregressive}. 
% \newedit{Edit this part later to add VQA module}Specifically, the NPA identifies multiple combination patterns from a basket set by looking up pattern codebooks. 
% Each of the selected pattern representations is then used to pick out contributing products of it from the basket and construct corresponding context for the incomplete combination to spot the missing components and support within-basket recommendation recommendation. 
% Section~\ref{sec:vqa} describes
% More specifically, 
To explicitly model combination patterns, the NPA framework comprises a network of Vector Quantized Attention (VQA) modules (detailly discussed in Section~\ref{sec:vqa}), which identify potential item \emph{combination
patterns} via lookup in a trainable combination pattern codebook, then infer a basket's \textit{context} to 
predict the missing components of the identified pattern. 
To further capture multiple granularity levels and the aspects of combination patterns,
the NPA framework is equipped with multiple channels and multiple layers of VQA modules (similar to the design of Transformer models~\cite{vaswani2017attention, devlin2018bert, radford2018improving}). 
% Thanks to the nature of the attention mechanism, 
\newedit{Together with the Any-Order Autoregressive Training~\cite{uria2014deep, strauss2021arbitrary, hoogeboom2021autoregressive} method,}
the resulting NPA model (see Section~\ref{sec:MMM_NPA}) achieves weak order sensitive, addressing the research challenges of modeling interleaving behavior of users during a shopping session.
% supports multi-combination-pattern, multi-granularity \textcolor{red}{WBR}
% within-basket recommendation 
%for a wide variety of shopping scenarios (e.g., data with or without temporal information)
% \tina{, we need multiple channels to capture different focuses/definitions of a combination pattern in a similar fashion to CNN or multi-head transformer models.}

% \tina{I think here we could add a little bit of motivation, to distinct this work as well, it is anonymous, weakly ordered, relation between combination pattern and context, .}

%In Section~\ref{sec:MMM_NPA}, we describe the multi-layer, multi-channel NPA model architecture in detail, where it replies on multiple stacked and parallel VQA modules (in similar fashion with the Transformer models~\cite{vaswani2017attention}). 

To show the effectiveness of the proposed NPA framework, we conducted various quantitative and qualitative evaluations (see Section~\ref{sec:experiments}) by comparing NPA with state-of-the-art baseline models on three e-commerce and entertainment datasets. 
The results show that the NPA model consistently outperforms the baseline models on all three datasets with 5\%-25\% performance improvement. Furthermore, the proposed model exhibits remarkable self-interpretability by tagging in-basket items as recommendation explanations without any post-processing interpretation efforts.

\section{Preliminary}
% \wuga{2022-04-17 2:26 PM ADT Wuga is here. The editing is much more work than I thought.. }
% \subsection{Notation}
% % \newedit{Modify notation after the symbols have been finalized.}

% Before proceeding, we define the notation used on this paper as follows:
% \begin{itemize}
%     \item $\mathbf{x}$: product vector with total length of M such that $\mathcal{X} = \{x_1, x_2 \cdots x_M\}$
%     \item $\mathbf{s}$: a basket as a product sequence, $\mathbf{s}_t \in \mathcal{X}$ represents the $t^{th}$'s product in the basket. 
%     % \kai{This is not $t$'s it should be $t^{th}$ instead.}
%     \item For a training dataset with multiple basket records $\bm{\mathcal{B}} = \{\cdots \mathcal{B}^{(o)} \cdots\}$ with respective cardinality $|T_{\mathcal{B}}^{(o)}|$ \tina{maybe not include in here, but this will be used in lit review, place it here for now.}
% \end{itemize}
\subsection{Problem Statement}
\label{sec:prob_state}

%\newedit{This part can start to be rephrase after intro}
%\tina{Add formal definitions for combination pattern and context}
Let $\mathcal{X} = \{\mathbf{x}_1,\mathbf{x}_2, \cdots, \mathbf{x}_m\}$ be a set of $m$ items. Given $\mathcal{B}\subset \mathcal{X}$, an incomplete shopping basket, the goal of within-basket recommendation is to recommend the top-k products $\mathcal{R}=\{\mathbf{x}| \mathbf{x}\in \mathcal{X}, \mathbf{x}\notin \mathcal{B}\}$ 
%\tina{$=$ or $\in$ ?} 
such that $\mathcal{R}\cup \mathcal{B}$ maintains the shopping intention %\tina{$\mathcal{Z}$ ?} 
of original basket $\mathcal{B}$. As a shopping intention is an abstract concept, we alternatively implement it as a common item combination in historical data $\mathbf{z}\in Z$, where $Z = [\cdots \mathbf{z}_i \cdots]$ denotes all the potential statistically significant combination patterns. %\tina{$z$ is coming out of no where, need to add its mathematical expressions in the previous sentence, also $z$ might need subscripts and $\mathcal{Z} = \{z_1, z_2 ..., z_N\}$ to express its non-singularity.}. 
%A candidate from the recommendation set $\mathcal{R}$ is evaluated for 

Conceptionally, the suitability of a recommendation $\mathbf{x}\in\mathcal{R}$ should align with the following guidelines:
\begin{enumerate}
    \item It must match at least one item-combination pattern of the given basket.
    \item It has higher recommendation priority if it fits a combination pattern with a higher likelihood. 
    % given observed products in the basket.
    %\item If a candidate product fits a combination pattern that has higher likelihood given observed product in basket, it has higher recommendation priority. 
    %\tina{Given the observed products in a basket, the recommendation priorities or probabilities of candidate products corresponds to the likelihood of the combination pattern it fits to or matches.}
    \item It has a higher recommendation probability if it fits multiple combination patterns.
\end{enumerate}

In this work, we do not intend to conflate the personalization task~\cite{ariannezhad2023personalized, liu2020basket, liu2020basconv} with the basket completion task but leave it as a downstream task, considering multiple realistic factors in real-world settings, such as 1) the sparsity of the basket data for personalization, 2) well-established post personalization strategy/pipeline, 3) a large number of anonymous users.
% For personalized complementary recommendation, in addition to the conditions above, the ideal candidate product should also match user preferences. \kai{Are we going to mention personalization here?}\tina{I think better not to..., let's see what wuga says}

\subsection{Related Work}
In this section, we summarize previous efforts in the field of within-basket recommendation and preliminary knowledge that significantly inspired this work.

% \wuga{2022-04-17 3:45 PM ADT Wuga is here.  }

% : Within-basket Recommendation, the Attention Mechanism, and the Any-Order Autoregressive Models. 
% \begin{itemize}
    % \item Previous use and terms for context, reason for its need and multi-context. \tina{may not need}
    % \item Sensitive to outliers - Why VQA \tina{may not need}
    % \item Order sensitivity, discussion. 
    % \item Use Hypergraph~\cite{feng2019hypergraph} to perform \tina{this is actually a different problem - next basket recommendation, not really in the same scoe.}
    % \item Set expansion (SE) refers to expanding a given partial set of objects into a more complete set~\cite{wang2008automatic}.
% \end{itemize}
% \tina{Several product representation learning methods are based on the skip-gram framework [22]. Essentially, they seek item representations which are useful for predicting contextual (related) items or users, by defining different ``context windows.", GNN-based models typically do not rely on the item add-to-cart sequence.}
\subsubsection{\bf Within-basket Recommendation}
\label{sec:lit_review_wbr}
 As within-basket recommendation (WBR) aims to complete a basket during a user's shopping session, an essential factor to model is item-to-item associations. 
% therefore identifying complements based on frequent co-purchases is the first rule of thumb for the vast majority of CPR research papers\cite{yu2019complementary}.
\newedit{Association rule learning~\cite{agrawal1994fast, tan2004selecting, gouda2001efficiently}, as a classic solution, explores relationships among co-purchased products.
% leading to remarkable success in the WBR research history. 
However, they do not scale well with the increasingly large transaction datasets within the e-commerce domain~\cite{yu2019complementary}.}
\newedit{Conventional latent representation approaches such as Item-item Collaborative Filtering (CF)~\cite{linden2003amazon}, and VAE-CF~\cite{liang2018variational} could also apply to WBR. Since those algorithms were not designed for WBR, their performances are often suboptimal. With the emergence of neural representation learning methods~\cite{mikolov2013efficient, mikolov2013distributed, pennington2014glove}, later ML-based models~\cite{grbovic2015commerce, barkan2016item2vec, vasile2016meta, wan2018representing} addressed various modeling issues through product representation learning, yielding reasonable WBR performance.} 

\newedit{
Although many models~\cite{le2017basket, wan2018representing, yu2019complementary, xu2020knowledge, liu2020basconv, li2022modeling, ariannezhad2023personalized} were proposed to address different aspects of the basket completion task, they often do not explicitly handle the co-existence of 1) multiple intents, 2) multiple granularities of intents, and 3) interleaving intents in a shopping session.}
\newedit{
In fact, \textit{Multi-combination Pattern} is a commonly studied topic for recommender systems, where it focuses on modeling multiple user intentions (e.g., food and toiletries) reflected by the shopping basket. Previous studies~\cite{liu2020basket, shen2023temporal, wang2021learning, chen2022intent}} 
% in recommendation areas
\newedit{ typically utilize Graph Neural Networks (GNN) to discover and model multiple intents based on sessional or sequential user interaction data, while other neural network frameworks~\cite{cen2020controllable, chen2021multi, li2023gpt4rec} have been proposed to address similar issues. \textit{Multiple Granularity Levels of Combination Patterns}, on the other hand, refers to different levels of characteristics of the combination patterns (e.g., specific interests in sweet flavors or general interests in desserts for the combination pattern of gelato).
% diverse user interests leading to the user-item interactions. (e.g., purchasing bananas for making breakfast vs. a smoothie). 
Unfortunately, the latter topic received relatively less attention in the existing literature~\cite{guo2022learning, song2021capturing} within the field of WBR, not to mention jointly modeling factors of multiple intents and their multi-granularity levels along with users' interleaving shopping behaviors.
} This research gap motivated our work. 

\begin{figure*}[t]
    \centering
    \begin{subfigure}{0.49\linewidth}
    \centering
        \includegraphics[width=0.95\textwidth]{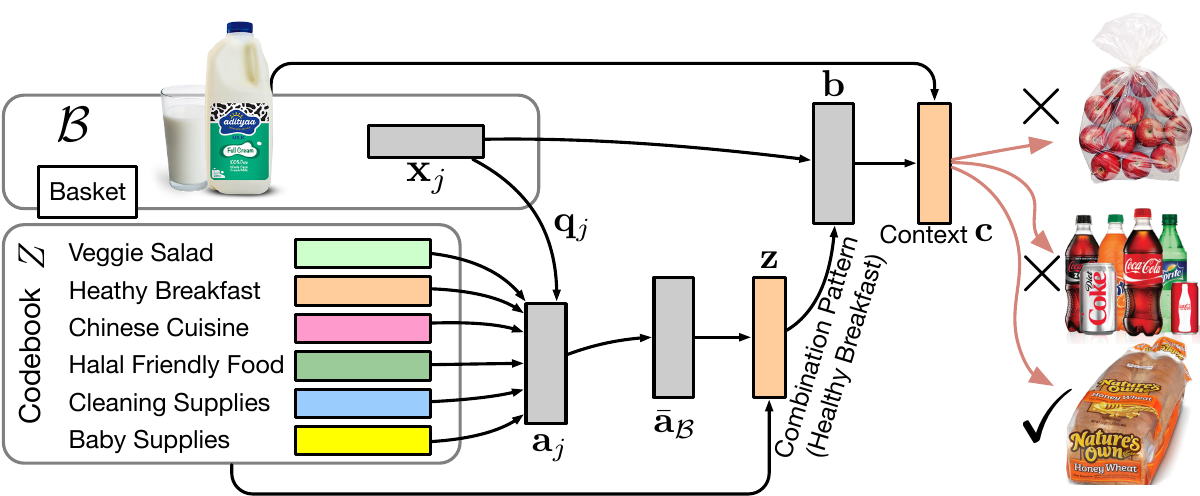}
    \caption{Basket with Single Product}
    \end{subfigure}
    \begin{subfigure}{0.49\linewidth}
    \centering
        \includegraphics[width=0.95\textwidth]{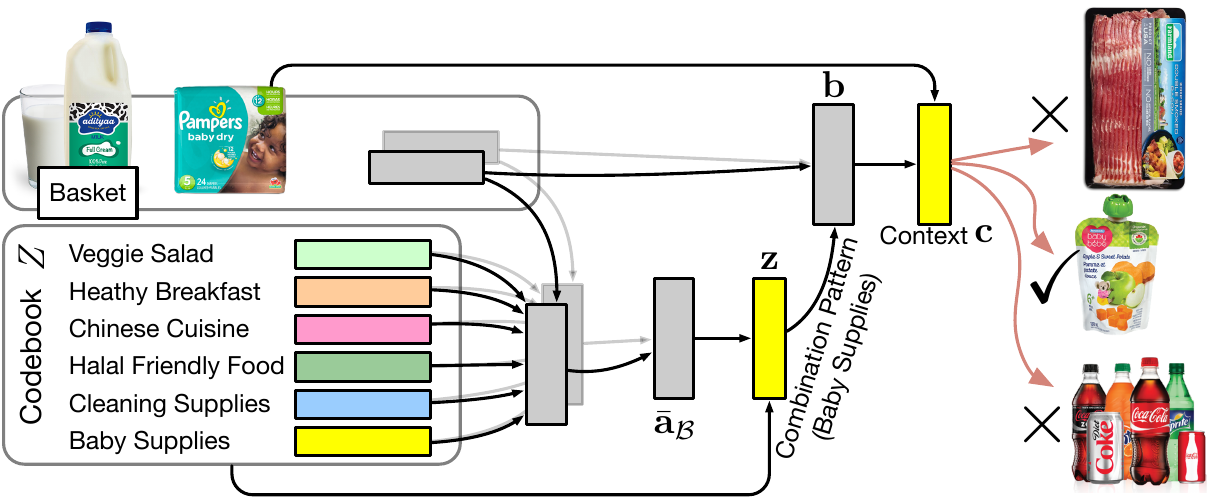}
    \caption{Basket with Multiple Products}
    \end{subfigure}
    \caption{Demonstrative figures of Vector Quantized Attention (VQA) module with two examples. Given products in the basket and a trainable codebook of combination patterns, VQA module estimates the basket's context by leveraging a pair of attention components. With more products included in the basket, the combination pattern estimated changes over time.}
\label{fig:vqa}
\vspace{-3mm}
\end{figure*}  

\subsubsection{\bf User Interest Discovery with Attention}
\newedit{
In recommender system research literature, the attention mechanism is widely adopted to model dynamic user interests.
% both the multi-CP
% ~\cite{cen2020controllable, chen2021multi, wang2021learning, shen2023temporal} 
% and multi-granularity of CP problems
% ~\cite{guo2022learning, song2021capturing}. 
ComiRec~\cite{cen2020controllable} applies the self-attentive method to its multi-interest extraction module. Similarly, MDSR~\cite{chen2021multi} applies a multi-head attention mechanism to model latent user interest. 
% MSGIFSR~\cite{guo2022learning} models a heterogeneous graph attention network to get the representations for intent units from all granularity levels.
% The main task of sequential recommendation systems is to predict the continuous items that the users may interact with given their historical interaction behavior, typically on the modern e-commerce platform.
% While the studies mentioned above employ the attention mechanism for estimating combination patterns, attention-based models~\cite{li2017neural, kang2018self, sun2019bert4rec, li2023gpt4rec} have demonstrated their effectiveness in capturing the context of users' activities. 
The attention mechanism is also commonly used to predict users' continuous interactions with the candidate items based on their historical behaviors on modern e-commerce platforms.
SASRec~\cite{kang2018self} leverages a self-attention module to capture long-term semantics using the attention mechanism. In a similar fashion, BERT4Rec~\cite{sun2019bert4rec} further improves the performance by a bidirectional architecture. 
% A recent concurrent work GPT4Rec~\cite{li2023gpt4rec} adapts GPT-2 language model~\cite{wolf2020transformers} to create a generative framework for personalized recommendation. 
% Within the era of sequential recommendations, the session-based recommendation is a field analogously similar to the problem we study in this work, where the users are anonymous in each shopping session. 
These works utilize attention-based models for either combination pattern estimations or context estimations for better recommendation performance. However, the research gap in modeling and distinguishing both the concept of multi-granular multi-combination patterns and the basket context remains an important yet less focused challenge for WBR. 
% Despite the achievements of previous works, we noted some persisting research gaps in the WBR field, which motivated the work we present here. 
% In addition to the outlined limitations above, we note that in real-world e-commerce platforms or WBR applications, the order in which users place items in the basket is neither strictly sequentially restricted~\cite{kang2018self, sun2019bert4rec, liu2020basket, ariannezhad2023personalized}, nor completely random~\cite{hu2017diversifying, wang2017perceiving}, but rather weak ordering(i.e., assuming items in basket is selected without an explicit order, useful ordering information may be captured by the model.) \tina{change to more proper statements}. 
%These limitations motivated the work we present here.
}

\section{Vector Quantized Attention (VQA)}
\label{sec:vqa}
% \subsubsection{Maximum A Posterior Estimation}
% \subsubsection{Bayesian Model Averaging for Multi-context Handling}
% \subsection{Combination Pattern Recognition with Set Attention Module}
% We start with transforming complementary recommendation task into a probabilistic inference task. Conceptually, 
%\tina{Explain what is vector quantized, and why at the beginning.}
% \kai{Insert: Now ..}
In this section, we present the Vector Quantized Attention~(VQA) module, the basic building block of NPA. 
%VQA is a univariate-conditional-modeling neural-network architecture that is specifically designed to support WBR. 
In particular, the VQA module models the single-item set-expansion task as a conditional distribution of $p(\mathbf{x}|\mathcal{B})$ in an order-agnostic manner~\footnote{While higher-level basket evolution over time (reflecting an evolution of shopping intentions) is important, one-step set expansion should remain item-order agnostic considering the frequent interleaving behavior of users during shopping sessions.}, where during the recommendation phase we seek the item $\mathbf{x}\in \mathcal{X}_{/\mathcal{B}}$ that maximizes the conditional likelihood, $\argmax_{\mathbf{x}} p(\mathbf{x}|\mathcal{B})$.

% In particular, with a non-empty basket $\mathcal{B}$ and the conditional distribution of $p(\mathcal{X}_{/\mathcal{B}}|\mathcal{B})$, complementary recommendation is essentially a search task, where we look for an item $x\in \mathcal{X}$ that maximizes the conditional likelihood such that $\argmax_x p(x|\mathcal{B})$. However, modeling the univariate conditional distribution $p(\mathcal{X}_{/\mathcal{B}}|\mathcal{B})$ is challenging; naively modeling it through neural networks would fail to accomplish the modeling expectations described in Section~\ref{sec:prob_state} due to lack of regularization. 

The VQA module explicitly models two concepts, {\em combination pattern} $\mathbf{z}$ and {\em context} $\mathbf{c}$, that serve as an inductive bias and allow the univariate conditional to be decomposed as follows 
% \kai{$C_i$ includes $B$?}
\begin{equation}
\begin{split}
p(\mathbf{x}|\mathcal{B}) 
&= \sum_i p(\mathbf{x}|\mathbf{c}_i
% , \mathcal{B}
)p(\mathbf{c}_i|\mathcal{B}) 
\\
&= E_{p_{\Theta}(\mathbf{c}_i|\mathcal{B})}[p_{\vartheta}(\mathbf{x}|\mathbf{c}_i)] 
\\
&= E_{p_{\theta}(\mathbf{z}_i|\mathcal{B})\delta[\mathbf{c}_i - f_\phi(\mathbf{z}_i, \mathcal{B})]}[p_{\vartheta}(\mathbf{x}|\mathbf{c}_i)],
\label{eq:simple_rec}
\end{split}
\end{equation}
where $\mathbf{z}_i$ 
%\kai{Just to confirm this $z$ is the same as the $z$ in Section 2.2 right?} 
represents a possible item-combination pattern of the basket $\mathcal{B}$, and $\mathbf{c}_i = f_\phi(\mathbf{z}_i, \mathcal{B})$ denotes a corresponding context. The context reflects the current status of the basket in terms of fulfilling a given combination pattern. Intuitively, a context carries information of {\em what is still missing for the basket to fulfill a combination pattern (intention)}. 
In fact, given an identified context $\mathbf{c}_i$, by recommender system conventions~\cite{wan2018representing, xu2020knowledge}, the probability of recommending a particular product $\mathbf{x}_j$ can be simply formalized as a softmax function over the dot product between the embeddings of context and candidate products as follows
\nopagebreak[1]
\begin{equation}
    p_{\vartheta}(\mathbf{x}_j|\mathbf{c}_i) = \frac{\exp(\mathbf{e}_j^{\top} \mathbf{c}_i)}{\sum_{j'}\exp(\mathbf{e}_{j'}^{\top} \mathbf{c}_i)},
\label{eq:single_score}
\end{equation}
%\nopagebreak[1]
where $\mathbf{e}$
% $\mathbf{e}_j$ 
% \kai{insert: and $\mathbf{e}_j'$ or just say $\mathbf{e}$}
denotes trainable product embeddings. Hence, the modeling challenge is reduced to estimate informative context given a basket such that 
$p(\mathbf{c}_i|\mathcal{B}) = p_{\theta}(\mathbf{z}_i|\mathcal{B})\delta[\mathbf{c}_i - f_\phi(\mathbf{z}_i, \mathcal{B})]$, where $\delta$ is a delta function that returns 1 when $\mathbf{c}_i$ equal to $f_\phi(\mathbf{z}_i, \mathcal{B})$. Figure~\ref{fig:vqa} illustrates the computation flow of the VQA module.
% broken down into three relatively independent sub-tasks, namely 1) combination pattern identification $p_{\theta}(z_i|\mathcal{B})$, 2) context estimation $\mathbf{c}_i =\phi(z_i, \mathcal{B})$, and 3) next item recommendation given a combination pattern and products already in the basket $p_{\vartheta}(x|\mathbf{c}_i))$. 

% In the following sections, we describe how we model $p(\mathbf{c}_i|\mathcal{B})$ with novel Vector Quantized Attention~(VQA) Module by further decomposing it into two components, namely 1) combination pattern identification $p_{\theta}(\mathbf{z}_i|\mathcal{B})$, and 2) context estimation $\mathbf{c}_i =\phi(\mathbf{z}_i, \mathcal{B})$.

% \wuga{2022-04-17 8:57 PM ADT. I reached this point, but above section needs further re-wording. The writing quality is way to rough. Well, I wrote this long ago I know.}
\subsection{Combination Pattern Identification}
\label{sec:simple_combination_pattern}
%\tina{The deterministic mapping is outlier sensitive, that is why we use quantized vector to mitigate effects of large variance}
The most straightforward technique of inferring a combination pattern is to learn the deterministic mapping between basket $\mathcal{B}$ and each possible combination pattern $\mathbf{z}_i$ through universal function estimators (neural networks) such that $p(\mathbf{z}_i| \mathcal{B}) = f_{\theta}(\mathcal{B})_i$, where $f_{\theta}(\mathcal{B})_i$ denotes the output index $i$ of function $f_\theta$ that corresponds to the probability of the combination pattern $\mathbf{z}_i$. 
%However, this approach has two limitations. 1) Since the input feature dimension of a neural network is often fixed and ordered, it conflicts with our intention of supporting flexible number of products in a basket (which is a set). While binary encoding of presence of products in the basket is legit alternative, it negatively affects model generalization. 2) 
However, since products in baskets are often incomplete and noisy in terms of satisfying a coherent combination, inferring \newedit{a combination pattern on such baskets} without proper heuristics 
% a combination pattern on such baskets 
can result in large predictive fluctuation that greatly undermines inference performance. 

In this work, we propose a more stable inference logic that leverages the neural attention mechanism and vector quantization~(VQ) learning~\cite{van2017neural}.
%Recall that a codebook is commonly referred as a shared, trainable, hidden embedding list. 
% Here, we follow the general definition of codebook and control the size of the codebook with one hyperparameter. The value of this hyperparameter should be approximately the number of possible combination patterns in a dataset. 
%In the following description, we use $Z$ to denote the combination pattern codebook.
Concretely, for each item $\mathbf{x}_j\in \mathcal{B}$, based on the classic self-attention formula, we first generate its query, key, and value vectors through linear projections
\begin{equation}
\mathbf{q}_j = W_q\mathbf{x}_j, \quad \mathbf{k}_j = W_k\mathbf{x}_j, \quad \text{and }\mathbf{v}_j = W_v\mathbf{x}_j,
\label{eq:product_qkv}
\end{equation}
where we have re-used $\mathbf{x}_j$ to represent (trainable) item features of item $\mathbf{x}_j$, in order to avoid introducing additional notation.
%For now, we focus on the query vector $\mathbf{q}_j$ that we use to search for relevant combination patterns from a codebook. 

% A codebook is commonly referred as a shared, trainable, hidden embedding list. Here, we follow the general definition of codebook and control the size of the codebook with one hyperparameter. The value of this hyperparameter should be approximately the number of possible combination patterns in a dataset. In the following description, we use $Z$ to denote the combination pattern codebook.
%(as opposite to the set representation $\mathcal{Z}$, where $\mathbf{z}_i$ denotes the quantized representation of combination pattern $z_i$.
% \kai{Just double check. Is $Z$ the same as the definition in Section 2.2?}
% \kai{I am a bit confused between $\mathbf{z}_i$ and $z_i$ defined here and Section 2.2}.

Next, we infer the keys $\tilde{K}$ of candidate combination patterns by mapping each combination pattern $\mathbf{z}\in Z$ through $\tilde{\mathbf{k}}=W_{\tilde{k}} \mathbf{z}$. With these keys, the conditional distribution of the combination patterns given an item $\mathbf{x}_j$ can be modeled as attention
\begin{equation}
\mathbf{a}_j = p_\theta(Z|\mathbf{x}_j) = \exp\left(\frac{\tilde{K}\mathbf{q}_j}{\sqrt{|\mathbf{q}_j|}}\right) / \zeta, \quad
% \text{where partition function } \zeta = \sum_{i} \exp\left(\frac{\tilde{\mathbf{k}}_i^\top\mathbf{q}_{j}}{\sqrt{|\mathbf{q}_{j}|}}\right).
\end{equation}
% Note that, since the general assumption for products in a complete basket are order insensitive 
% %\tina{The default assumption of this work is that the products in a basket is order insensitive}\wuga{this is only true for VQA module, not for NPA as described in Section 4}
% , 
%where partition function $\zeta = \sum_{i} \exp\left(\frac{\tilde{\mathbf{k}}_i^\top\mathbf{q}_{j}}{\sqrt{|\mathbf{q}_{j}|}}\right)$. 
where $\zeta$ denotes a partition function that keeps $\sum_i a_{ji} = 1$.
Here the combination pattern $\mathbf{z}$ is represented as a trainable embedding vector, serving as VQ for later look-up during the inference phase.

To further obtain the overall combination pattern beliefs of a basket, we adopt a straightforward but reasonable solution -- averaging individual attention of all products in the basket as follows
\begin{equation}
\bar{\mathbf{a}}_{\mathcal{B}} = p_{\theta}(Z| \mathcal{B}) = E_{j} [\mathbf{a}_j], \quad \text{where } j\in \{j|\mathbf{x}_j\in \mathcal{B}\}.
\label{eq:avg_attention}
\end{equation}
% Indeed, this solution aligns with the order-invariant learning literature. 
%\wuga{2022-04-17 10:07 PM ADT. I reached this point. My brain is stop working.. I will continue later or tomorrow.}
% which represents the probability of a combination pattern given basket $p_{\theta}(Z| \mathcal{B}) = \bar{\mathbf{a}}_{\mathcal{B}}$.
% we desired. 

The categorical distribution $\bar{\mathbf{a}}_{\mathcal{B}}$ will be further used to extract the possible combination patterns embeddings for the basket, where we propose three strategies for different purposes:
\begin{itemize}
    \item Greedy Search: selects a combination pattern that has maximum attention values
    \begin{equation}
    \mathbf{z} = \mathbf{z}_i, \quad \text{s.t.} \quad i = \argmax \bar{\mathbf{a}}_\mathcal{B} \quad \text{and} \quad \mathbf{z}_i \in Z.
    \label{eq:z_max}
    \end{equation}
    \item Weighted Average: computes the weighted expectation of multiple combination patterns 
    \begin{equation}
    \mathbf{z} = Z^\top\bar{\mathbf{a}}_\mathcal{B}.
    \end{equation}
    \item Sampling: samples a combination pattern based on the conditional distribution $p_{\theta}(Z| \mathcal{B})$ 
    \begin{equation}
        \mathbf{z} \sim \textit{Categorical}(Z, \bar{\mathbf{a}}_\mathcal{B})
    \label{eq:z_sample}
    \end{equation}
\end{itemize}

Each strategy has a specific application scenario.
Greedy search is suitable for exploiting the most promising user intention for future recommendations. Weighted Average is valuable for summarizing multiple potential user intentions. Sampling is pivotal in supporting exploration, where it enhances the model's generalization and dynamic when users exhibit interleaving behavior during shopping journeys.
% Greedy search is suitable for creating a basic NPA model with a single-layer, single-context VQA module. Weighted Average is valuable for multi-layer NPA models, where it supports information extraction from lower layers. Sampling is pivotal in supporting diverse recommendation, where it enhances the model's generalization when users exhibit interleaving behavior during shopping journeys.
% Greedy search is particularly useful for building a simple NPA model that only contains a single layer, single context VQA module. 
% Weighted Average is helpful in the multi-layer NPA model setting, where it is used in the lower layers to support information extraction. 
% Sampling plays an important role in the multi-context setting, where we expect the full functional NPA model to produce multiple contexts for a basket such that the model can generalize better when the users show interleaving behavior during their shopping journeys. 
%We describe \textcolor{red}{context extraction} further in Section~\ref{sec:MMM_NPA}.

%Later, we will describe the details in which each of the strategies is applied. 
%All the parameters involved in the Vector Quantized Attention module are thereby $\theta = [W_q, W_{\tilde{k}}, Z]$.

% And, all the parameters involved in the VQ Attention module are thereby $\theta = [W_q, W_{\tilde{k}}, Z]$.

\subsection{Context Estimation}
\label{sec:simple_context_estimation}
Once a combination pattern of a basket is identified with the method described in Section~\ref{sec:simple_combination_pattern}, we can now estimate the context $\mathbf{c}$ of the basket $\mathcal{B}$ with respect to the combination pattern $\mathbf{z}$, such that $\mathbf{c} =f_\phi(z, \mathcal{B})$. 
% Note, 
% a context 
% % is a slightly different concept from a combination pattern; it 
% reflects the current status of the basket in terms of fulfilling a given combination pattern. Intuitively, a context carries information of {\em what is still missing for the basket to fulfill a combination pattern}. 
%And, obviously, not all products in the basket match the given combination pattern. 
%In Equation~\eqref{eq:simple_rec}, the context is represented as a function $\phi(z_i, \mathcal{B})$. 

% To model the context $\mathbf{c} =\phi(z, \mathcal{B})$, 
% % we again use the attention mechanism to pick out the products in the basket that contribute to the identified combination patterns. 
% \textcolor{red}{we employ an attention mechanism to identify the products within the basket that contribute to the identified combination patterns.}
Specifically, we first transform the extracted combination pattern embedding $\mathbf{z}$ (from Equation~\eqref{eq:z_max}-\eqref{eq:z_sample}) into a context query
\begin{equation}
    \varrho = W_\varrho \mathbf{z},
\label{eq:varrho}
\end{equation}
that serves to highlight the relevant items in the current basket. Then, we estimate the conditional distribution of items given the context query such that
\begin{equation}
\mathbf{b} = \exp\left(\frac{K\varrho}{\sqrt{|\varrho|}}\right) / \zeta \quad \zeta = \sum_j \exp \left(\frac{\mathbf{k}_j^\top\varrho}{\sqrt{|\varrho|}}\right),
\end{equation}
where item keys $K$ come from Equation~\eqref{eq:product_qkv}. With attention highlighted products, we model the context $\mathbf{c} $ through a linear combination
\begin{equation}
\mathbf{c} = f_\phi(\mathbf{z}, \mathcal{B}) = V^\top\mathbf{b},
\label{eq:context}
\end{equation}
where $V$ is obtained from Equation~\eqref{eq:product_qkv}, representing individual context contribution $\mathbf{v}_j$ of each item $\mathbf{x}_j$. 
% Obviously, 
% A context is always correlated with a combination pattern; there will be multiple contexts inferred through Equation~\eqref{eq:varrho}-\eqref{eq:context} if we sample multiple combination patterns. 
% Since attention $\mathbf{b}$ extracts context information from the VQA module inputs, it can also serve to explain the recommendation given by the NPA model, as demonstrated in Section~\ref{sec:recommendation_explanation_via_item_attention}.
%as we will discuss the model self-interpretability in Section~\ref{sec:prediction_explanation_context_attention} \tina{find this section}.

%It is easy to note that, when a basket has only one combination pattern and its attention $\mathbf{b}$ is an uniform distribution over all products in the basket, the context estimation falls back to prod2vec-like models where basket context is an average of all individual product contexts such that $\mathbf{c} = E_j[\mathbf{v}_j]$, where $j\in \{j | x_j \in \mathcal{B}\}$. 

By combining descriptions in Section~\ref{sec:simple_combination_pattern} and \ref{sec:simple_context_estimation}, we can now estimate the probability of the context for a given basket $p_\Theta(\mathbf{c}|\mathcal{B})$ as desired in Equation~\eqref{eq:simple_rec}. The parameter group involved in VQA Module are therefore $\Theta = [\theta, \phi] =  \left[[W_q, W_{\tilde{k}}, Z],[W_k, W_v, W_{\varrho}]\right]$, resulting in a remarkably simple yet efficient computation unit.

VQA is the base computation unit of NPA model. In fact, VQA serves NPA in a similar way that the Self-attention mechanism serves Transformer~\cite{vaswani2017attention}, where multi-head, multi-layer extensions are immediately accessible, as we will describe next.
%In the next section, we describe how to train a base Neural Pattern Associator model.

\section{Neural Pattern Associator (NPA)}
\label{sec:MMM_NPA}
\begin{figure}[t]
    \centering
        \includegraphics[width=0.95\linewidth]{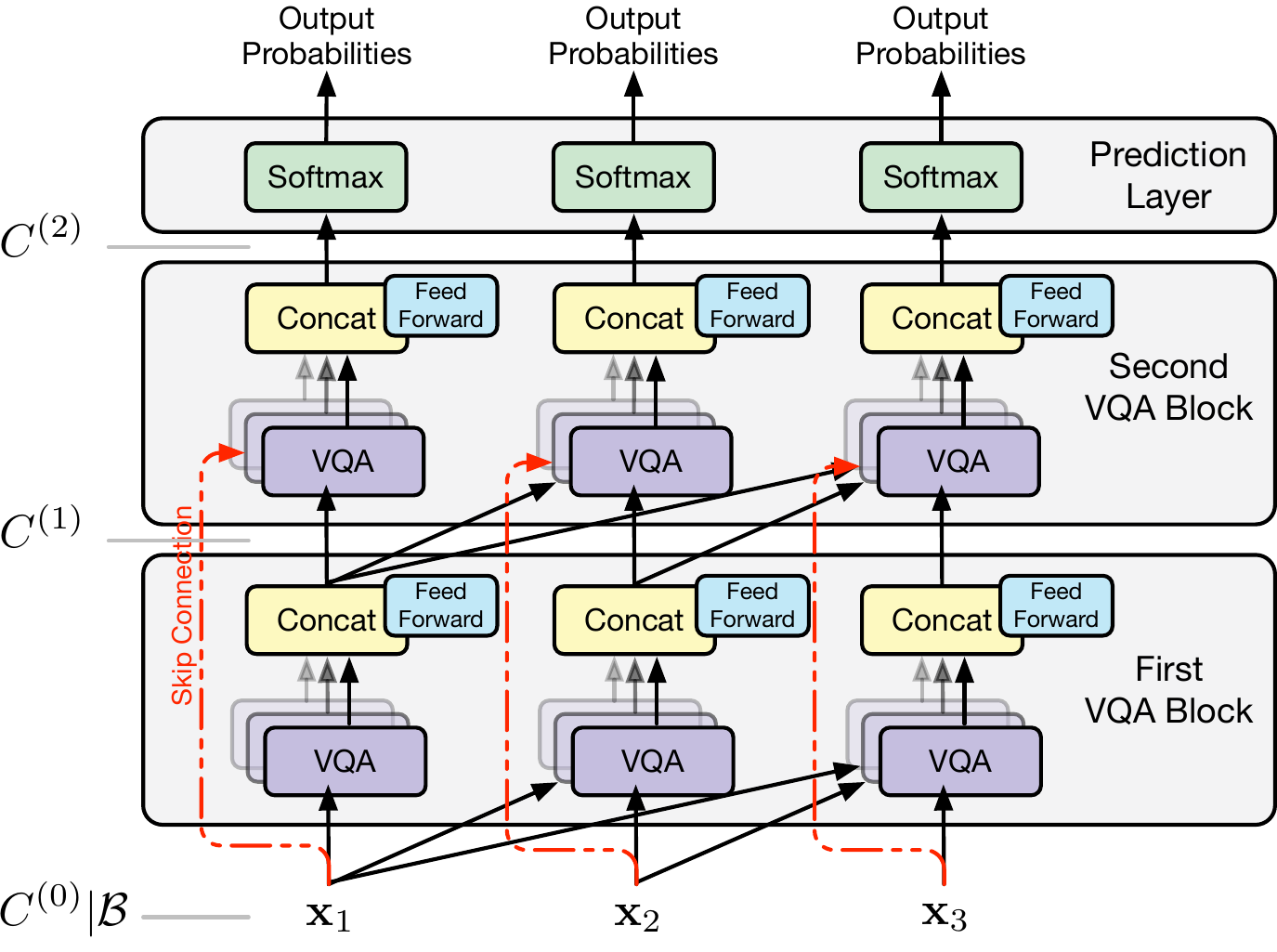}
    \caption{The high-level overview of Neural Pattern Associator model (NPA-SC). Figure shows two-layer NPA with three channels, where multiple VQA computation units are stacked in a similar fashion of multi-head attention. }
\label{fig:npa}
\vspace{-2mm}
\end{figure}

Now, we introduce Neural Pattern Associator~(NPA), a multi-layer multi-channel model consisting of multiple stacked and parallel VQA modules. A two-layer, three-channel NPA model is illustrated in Figure~\ref{fig:npa}. 

\subsection{Multi-layer Multi-step Context Extraction}
% \subsection{Multi-layer Multi-Step Context Extraction}
% \label{sec:shared_multi_context}
% Combination Pattern may have concept hierarchy such that smaller combination patterns are under the umbrella of a large combination pattern. E.g., Christmas Eve shopping cart may include Christmas decoration, Christmas dinner, and Christmas gift. To model the hierarchy of the combination patterns, we propose to use multi-layer paired attention model. 
Shopping intentions are multi-granular, where higher-level context information is often too abstract to be modeled with a single unit of VQA module.
% , so single-layer context extraction techniques struggle to model it precisely. 
% To extend the modeling capacity, 
% we therefore stack 
Stacking layers of VQA modules enables the NPA framework to gradually capture information through successive filtering. 
Specifically, with multiple steps of basket evolution over time $[1, \cdots t, \cdots T]$, the outputs of a single VQA module $\mathbf{c}_t = f_{\Theta}(\mathcal{B}_t)$ can be viewed as input sequence to the next layer:
% , with the following inference chain:
\begin{equation}
\mathbf{c}_t^{(l+1)} = f_{\Theta}^{(l+1)}\left(\hat{C}^{(l)}_{:t}\right), \, \text{s.t.} \,\, \hat{C}^{(l)}_{:t} = C^{(l)}_{:t} + C^{(l-1)}_{:t},
\end{equation}
 where $l$ denotes the layer index,  $ C^{(l)}_{:t}$ denotes all the estimated contexts from the lower layer $l$ up to step $t$ such that
 \begin{equation}
 C^{(l)}_{:t} = [\mathbf{c}_1^{(l)},\cdots\mathbf{c}_t^{(l)}].
 \end{equation} 
 In particular, $C^{(0)}_{:t}$ is the raw features of items in the basket $\mathcal{B}$ such that $C^{(0)}_{:t} = \mathcal{B}_t$. 
 % We also utilize residual connections to stabilize model training. 
 The non-last-layer VQA modules act as information filters (or encoders) and are not responsible for producing predictive contexts to support recommendations. Hence, to better preserve the information flow, we use the Weighted Average strategy described in Section~\ref{sec:simple_combination_pattern} to extract lower-level context information.

While modeling $p(\mathbf{x}|\mathcal{B})$ with a VQA module is agnostic to the item sequences in the basket, the multi-layer extension of NPA is an order-aware model due the side effect of taking lower-layer contexts as inputs (the order of basket context evolution) of the high-layer VQA modules. 
%\tina{do we need a separate paragraph below?}
% wuga: lets change it back if we don't have much space in the end
% As the NPA model is not order agnostic, in the following description, we use a particular permutation of basket $\mathbf{s}\sim \text{Perm}(\mathcal{B})$ to continue our discussion. 

The fundamental difference between NPA and Tranformer-based sequential models (e.g. SASRec~\cite{kang2018self}) lies in NPA's weak sensitivity to order.
% is that the NPA model is weak order sensitive; 
% the NPA model is sheltered from the risk of withholding strong position (order) dependency assumption, 
NPA avoids assuming strong position dependency by carrying a set of mutual independent codebook entries\footnote{At each time step $t$, Transformer uses the input of time step $t$ to generate the query to aggregate information, whereas NPA conducts votes among all inputs up to time step $t$ to look up a query (from a codebook) to aggregate information.}, addressing the problem of multiple interleaving intentions. In practice, when data lacks order information, 
% multiple 
% random data permutations during NPA model training 
training NPA model with random data permutations
emulates the behavior of order-agnostic models like AO-ARMs~\cite{uria2014deep, strauss2021arbitrary, hoogeboom2021autoregressive}\footnote{Detailed formulation of such training objective is presented in Appendix~\ref{app:training_obj_wo_temporal_info}}. 
% We provide more detailed formulation of such model in Appendix~\ref{app:NPA_multiLayer_multiStep}.
% (see Section~\ref{sec:ao_arm}).

% By stacking multiple layers of VQA modules, the non-last layer VQA modules are no longer responsible for producing predictive contexts to support the next product recommendation. Rather, they act as information filters (or encoders). Hence, to better preserve the information flow, we use the Weighted Average strategy in VQA module described in Section~\ref{sec:simple_combination_pattern} to extract lower-level context information.

% \subsection{Multi-channel Context Extraction}
\subsection{Multi-channel Context Extraction}
\label{sec:multi_channel_cp_extraction}
% A complex shopping combo may be consist of multiple aspects of focuses or smaller granularity of intentions. 
% E.g., A shopping cart indicating Italian dietary preference could include different properties such as flavor (spicy, salty, or sour) or different formality (dinner, snack). 
Correctly estimating a shopping basket's overall context may need evidence support from various focal aspects. 
% involves evaluating multiple aspects of focus. 
For example, recognizing a shopping basket serves for Christmas celebrations could rely on multiple evidences, such as themed tableware, decorations, and ingredients for roasting turkey, etc. 
To explicitly capture those decomposable aspects of the context of a basket, we incorporate multiple context channels, similar to convolutional or multi-head self-attention layers. 
Concretely, the multi-channel extension of NPA model maintains multiple parallel-running VQA modules in each layer, where we collect the estimated context with the linear projection of their concatenation\footnote{We omit the time step index $t$ and use the first layer of NPA model in the above equation to maintain clarity.}
\begin{equation}
\begin{aligned}
   \!\!\!\! \mathbf{c} \!=\! W_\sigma\mathbf{c}'\!, \,
   \mathbf{c}' \!=\! \textit{Concatenate}([\cdots \! \mathbf{c}^{h} \! \cdots]) , \, 
   % &\text{and}
   \mathbf{c}^{h} \!=\! f_{\Theta^{h}}(\mathcal{B}) ,
\end{aligned}
\label{eq:lower_level} 
\end{equation}
where $h$ denotes the channel index, $\mathbf{c}^{h}$ is an aspect of context, and $W_\sigma$ is a linear projection matrix. Note that each channel maintains an independent set of parameters $\Theta^h$.
% We further discuss the formulations in Appendix~\ref{app:multi_channel_context_extraction}.}
With Equation~\ref{eq:lower_level} , we propose the first version of NPA model with the squashed-context (SC) mechanism, where the setting in Equation~\ref{eq:lower_level} applies to all layers of the VQA modules.
% Here, we omit the time step index $t$ and use the first layer of NPA model in the above equation to maintain clarity.
% In addition, for the lower-level combination pattern extraction (in the sense of multi-layer extraction that we well describe later), enforcing single aspect property through sampling is unnecessary and can cause model training fluctuation. Hence, instead of sampling through Equation~\eqref{eq:sampled_combination_pattern}, we adopt greedy solution such that 
% \begin{equation}
% \mathbf{z}^h = Z_i^h, \quad \text{where } i = \argmax \bar{\mathbf{a}}_\mathcal{B}^h.
% \label{eq:max_combination_pattern}
% \end{equation}
% Its affect is similar to the max-pooling of a CNN layer, where the most significant features are selected for further processing.

While concatenating extracted contexts enables the NPA model to extract overall context and produce coherent recommendations, the sub-contexts of a basket could be diverse in practice. To capture the diversity of contexts, we propose an alternative context extraction mechanism -- multi-context (MC).
Specifically, instead of retaining channel-wise codebooks $Z^h\in \Theta^h$ in the last layer, NPA-MC\footnote{See Appendix~\ref{app:npa_mc_details} for illustration of the model architecture} uses a shared codebook $Z$ over all the VQA modules,
% in that layer}
% For the last layer of the extended multi-channel NPA model, however, while we still retain multiple parallel VQA modules, we do not retain channel-wise codebooks $\bm{Z} = \{\cdots Z^{h}\cdots\}$ as used in Equation~\eqref{eq:lower_level} but sharing a single codebook $Z$ over all VQA modules in that layer. 
% Correspondingly, we do not concatenate context as shown in Equation~\eqref{eq:lower_level}, but 
and treats each $\mathbf{c}^h$ as independent context estimation, resulting in multiple contexts $C_t = \{\cdots \mathbf{c}_t^{h}\cdots\}$ at each prediction step $t$. 
% that aim to diversify recommendations. 
To complement the diversified context extraction, we use the Gumbel sampling~\cite{jang2016categorical, maddison2016concrete} strategy (see  Equation~\eqref{eq:z_sample}) to sample different combination patterns that result in different contexts.
% \kai{Should we explicitly say ``gumbel sampling''?} 

% \kai{2023-04-16: My first pass for checking correctness is here.}
% \tina{2023-06-29: finished the above methodology section.}
\section{Training Objective of NPA}
\label{sec:VQA_module_training}
% In this section, we describe how to train the NPA model. Here, we consider two training scenarios, 1) Baskets contain temporal (or ordering) information when the products were put into the baskets, and 2) Baskets do not have temporal information associated with the products.
% \textcolor{red}{In this section, we describe the NPA model training objective.}

% \subsection{Data with Temporal Information}
% \label{sec:temporal_data}
For a training dataset containing temporal information,
% multiple 
basket records $\bm{\mathcal{B}} = \{\cdots \mathcal{B}^{(o)} \cdots\}$ have respective cardinality $|T_{\mathcal{B}}^{(o)}|$.
% if the temporal information 
% in which the products were put into the baskets 
% is available, 
A basket $\mathcal{B}^{(o)}$ can be represented as an item sequence $\mathbf{s}^{(o)}$, 
% we can represent a basket $\mathcal{B}^{(o)}$ as a product sequence $\mathbf{s}^{(o)}$
where $s_t^{(o)}\in \mathcal{X}$ represent the $t^{th}$
% \kai{$t^{th}$?} 
item in the basket. The item sequence enables autoregressive model training that  jointly maximizes the likelihood of an ensemble of univariate conditionals. Concretely, we maximize the lower-bound of observation likelihood as follows
\begin{equation}
    \argmax_{\Theta}\sum_{o}\sum_{t=2}^{|T_\mathcal{B}^{(o)}|} \log p_{\Theta}\left(s^{(o)}_{t}|\mathbf{s}^{(o)}_{<t}\right),
\end{equation}
where all individual conditional distributions $p_{\Theta}\left(s^{(o)}_{t}|\mathbf{s}^{(o)}_{<t}\right)$ are modeled 
% through the same NPA model 
with position masking. 
% In addition, 
We also explicitly encode the position information to capture temporal information; given item $s_t$, we retrieve its corresponding features $\mathbf{x}_j$ and override by $\hat{\mathbf{x}}_j = \mathbf{x}_j + \mathbf{p}_t$,
% \begin{equation}
%     \hat{\mathbf{x}}_j =  \mathbf{x}_j + \mathbf{p}_t,
% \end{equation}
where $\mathbf{p}_t$ is the 
% time-step (or position) 
position encoding. 

When training a NPA-MC model,
% is slightly different 
% from the method above. 
% since we encourage context diversity by explicitly modeling multiple contexts in the training objective,  
% always prefer to 
% we update the context that best fits the 
% training data \textcolor{red}{to encourage context diversity}, 
% observation
% leading to the following updated objective
% \kai{Is this equation too long? the reference number is below the equation.}
we only optimize the context that produces the highest observation likelihood to the ground-truth candidate product at each time step. 
\begin{equation*}
\label{eq:obj_no_temporal_info}
    \argmax_{\Theta} \! \sum_{o} \! \sum_{t=2}^{|T_\mathcal{B}^{(o)}|} \max_{\mathbf{c}_t^{(o)}\in C_t^{(o)}}\log p_{\vartheta}\left(s^{(o)}_{t}|\mathbf{c}_t^{(o)}\right) p_{\theta}\left(\mathbf{c}_t^{(o)}|\mathbf{s}^{(o)}_{<t}\right),
\end{equation*}
% we only optimize the context that produces the highest observation likelihood to the ground-truth candidate product at each time step. 

While the above loss functions are designed for datasets with temporal information, they could be adapted for datasets without temporal information, where positional embedding is removed. We provide extended formulations in Appendix~\ref{app:training_obj_wo_temporal_info}.

\section{Experiments}
This section evaluates the NPA model by comparing it to various baseline models on three industry-scale benchmark datasets, including one private industry dataset obtained from a large North American food retailer. We aim to answer the following questions:
\label{sec:experiments}
\begin{itemize}
    \item {\bf RQ1:} Is the proposed Neural Pattern Associator
    % ~(NPA) 
    model competitive in terms of within-basket recommendation performance in comparison to state-of-the-art models?
    \item {\bf RQ2:} How are the multiple shopping intentions in shopping baskets captured by the multi-channel NPA model?
    \item {\bf RQ3:} How sensitive is NPA's performance to different model settings, such as the embedding dimensions of combination patterns,
    number of VQA layers, and number of channels?
    % (attention heads), 
    % \item {\bf RQ3:} How interpretable is the NPA model regarding codebook encoding inspections?
    \item {\bf RQ4:} How interpretable is the NPA model regarding the prediction explanations?
    % \kai{I think this part is out of sync with the experiments below}
    % \item {\bf RQ4:} How do different variations of context estimation settings of NPA lead to different behaviors in terms of within-basket recommendations?
    % \item \tina{Add a plot for training curve between different number of VQA module layers}
\end{itemize}
% \tina{Plots make sure to have ratios of 5:2.5}
% Results are reproducible with link: 
% \url{https://shorturl.at/dyFOV}.
% \footnote{\url{https://shorturl.at/dyFOV}}

\subsection{Experiment Settings}
\subsubsection{Dataset.}
\label{sec:dataset}
We evaluate the within-basket recommendation performance on three real-world datasets for online grocery shopping and online entertainment, including  
Instacart\footnote{https://www.kaggle.com/competitions/instacart-market-basket-analysis/data}, 
Spotify\footnote{https://www.aicrowd.com/challenges/spotify-million-playlist-dataset-challenge}, 
% Tafeng\footnote{https://www.kaggle.com/chiranjivdas09/ta-feng-grocery-dataset},
and a private industry dataset\footnote{See Appendix~\ref{app:dataset_descriptions} for more information on the dataset descriptions.}.
% \footnote{Private industry dataset}.
% Table~\ref{tb:datasets} shows the statistics for the datasets after data preprocessing. 
% We further split each dataset --- specifically, leaving 20\% baskets as testing data and the other 20\% baskets as validation data. The remaining 60\% baskets are used to train the models. 
For each dataset basket, we follow a [0.6, 0.2, 0.2] train-validation-test split,
where the evaluation is conducted on baskets in the test sets with sampling. 
% For datasets with temporal information, we sample a consecutive sequence of items from each basket as the incomplete basket. In contrast, for datasets without temporal information, we randomly sample a set of items from each basket as the incomplete basket. Ground-truth labels come from the remaining items in the baskets that were not sampled as part of incomplete baskets.
Specifically, for datasets with temporal data (i.e., Spotify), we sample consecutive items from each basket as the model input while for datasets without temporal data (i.e., Instacart and the industry dataset), randomly selected items are the model input. Ground-truth labels are the remaining basket items.

\subsubsection{Baselines \& \newedit{Implementation Details.}}
For RQ1, 
we compare the proposed NPA model\footnote{See Appendix~\ref{app:model_implementations} for detailed model implementation descriptions.} with the following baseline models for the recommendation performance:
\begin{itemize}
\item {\bf Popularity (POP):} Recommend the most popular items to the global population is an intuitive baseline.
\item {\bf Co-purchase (CP):} Recommend items based on their co-purchase frequencies with the candidate items. 
% It often shows a strong performance when the dataset is small with minor noise.
\item {\bf Apriori~\cite{agrawal1994fast}:} One of the conventional frequent item set mining algorithms.
\item {\bf Item-item CF~\cite{linden2003amazon}:} Collaborative filtering is performed on the item-to-item co-purchase matrix. 
\item {\bf Prod2Vec~\cite{grbovic2015commerce}:} 
Word2Vec-like model learned through skip-gram training schema~\cite{mikolov2013distributed}. 
\item {\bf GraphSAGE~\cite{hamilton2017inductive}:} A graph framework that computes node representations in an inductive manner. 
% We utilize it on a basket-item interaction graph to 
which perform a link prediction task\footnote{In this graph, nodes represent items, and edges are present if both items appear in the same shopping basket.} on a basket-item interaction graph.
\item {\bf VAE-CF~\cite{liang2018variational}:} \newedit{A recommendation algorithm that uses variational autoencoder for collaborative filtering.}
\item {\bf SASRec\footnote{https://github.com/kang205/SASRec} ~\cite{kang2018self}:} 
A self-attention-based sequential recommendation model that captures long-term dependencies and
% A state-of-the-art 
% sequential recommendation model that leverages self-attention mechanisms to capture long-term dependencies and 
makes predictions based on a few relevant actions. 
% \kai{Add citation and description.}
%Given a purchased product, the method calculates cosine similarities to all other products in the vocabulary and recommends the top K most similar products. 
% \kai{I am not sure if this is the correct description for Prod2Vec. This seems to be the right description for Item-item CF.} \tina{updated}
% \item {\bf Sceptre~\cite{mcauley2015inferring}:} 
% Utilizes topic modeling on item textual features from review text and logistic regression for substitute/complement classification. Category information is also applied with a sparse encoding technique.
% (The topic modeling in the original paper can be used as aisle or department modeling for the Instacart dataset we are using) \kai{Need to be removed or replaced with another baseline.}
\item {\bf BERT4Rec\footnote{https://github.com/FeiSun/BERT4Rec}~\cite{sun2019bert4rec}: }\newedit{A state-of-the-art recommendation model that employs the deep bidirectional self-attention to model user behaviors sequences.}
% \item {\bf \newedit{GPT4Rec~\cite{li2023gpt4rec}: }} A concurrent work that uses the GPT~\cite{radford2018improving} architecture for recommendations. 
% \kai{Cite GPT4Rec. Note concurrent work}
% \item {\bf NPA-SingleContext (NPA-SC):} 
% \item {\bf NPA-MultiContext-item-input-only (NPA-SC-itemonly):}
% \item {\bf NPA-MultiContext (NPA-MC):} 
\end{itemize}

\begin{table*}[t!]
\caption{Within-basket recommendation results of the datasets. We omit the error bars as the confidence interval is in the 4th digit. The best-performing model for each evaluation metric is marked in bold, and the second best-performing model is underlined.
% ``DNF'' indicates a method did not finish running in 24 hours using a NVIDIA A100 GPU. \kai{Remove DNF}
% \tina{Explain the use of residual on the first layer. Spotify not using the residual} 
% \kai{I don't think we have to talk about the residual at the first layer. Our models overfit Spotify. To save cost, we decided not to re-train our models on Spotify. Based on the observations on the other three datasets and the performance of GPT improves after adding residual at the first layer, our models should also benefit from it.}\kai{I think we should consider using underline to indicate the second best performing. Our models are the best and second best performing models in most cases. This would also help to demonstrate that NPA-MC isn't that bad in TaFeng. Its performance is the second best across all metrics except for R@1 which its the best.}
}

\vspace{-2mm}
% \rowcolors{2}{gray!10}{white}
\resizebox{\textwidth}{!}{%
\begin{tabular}{c|ccccccccccccccc}
 
 \toprule
 
&Candidate Model&P@1&P@5&P@10&P@15&P@20&R@1&R@5&R@10&R@15&R@20&R-Precision&NDCG\\
\midrule
\multirow{11}{*}{\rotatebox[origin=c]{90}{Instacart}}&POP&0.0605&0.0419&0.0315&0.0261&0.0226&0.0153&0.0513&0.0752&0.0924&0.1061&0.0407&0.0729 \\

&CP&0.0761&0.0482&0.0362&0.0301&0.0263&0.0197&0.0601&0.0889&0.1099&0.1271&0.0473&0.0865 \\

&Apriori&0.0604&0.042&0.0331&0.0275&0.0237&0.0151&0.0512&0.0798&0.0985&0.1123&0.0404&0.0748 \\

&Item-item CF&0.0632&0.0447&0.0354&0.0301&0.0266&0.0176&0.0595&0.0918&0.1154&0.1345&0.0449&0.086 \\

&Prod2Vec&0.0227&0.0176&0.0141&0.0121&0.0108&0.0064&0.0242&0.0383&0.0489&0.0577&0.0174&0.0351 \\

&\newedit{GraphSAGE}&0.0233&0.0148&0.0109&0.009&0.0078&0.0066&0.02&0.0292&0.0361&0.0417&0.015&0.0278\\

&\newedit{VAE-CF}&0.0795&0.0499&0.038&0.0319&0.0281&0.0211&0.0633&0.0949&0.1188&0.1386&0.0496&0.0924 \\

&SASRec&0.0611&0.0426&0.0323&0.0269&0.0235&0.0156&0.052&0.0779&0.0966&0.1119&0.041&0.0749\\

% &\newedit{GPT4Rec}&0.0972&0.0636&0.0481&0.04&0.0348&0.0277&0.0861&0.1274&0.1569&0.1805&0.0645&0.1197 \\
&\newedit{BERT4Rec}&0.0599&0.0403&0.031&0.0257&0.022&0.0152&0.0491&0.0746&0.0913&0.1041&0.0393&0.0712 \\
% &\newedit{\textcolor{red}{ComiRec}}&\textcolor{red}{0.127}&\textcolor{red}{0.0404}&\textcolor{red}{0.0368}&\textcolor{red}{0.0308}&\textcolor{red}{0.028}&\textcolor{red}{0.0182}&\textcolor{red}{0.0284}&\textcolor{red}{0.0515}&\textcolor{red}{0.0643}&\textcolor{red}{0.0781}&\textcolor{red}{0.0387}&\textcolor{red}{0.0676} \\
\cmidrule{2-14}
&NPA-SC &{\bf0.0985}&\underline{0.064}&\underline{0.0485}&\underline{0.0404}&\underline{0.0351}&\underline{0.0279}&\underline{0.0864}&\underline{0.1281}&\underline{0.1577}&\underline{0.1815}&\underline{0.0648}&\underline{0.1204}\\

&NPA-MC &\underline{0.0973}&{\bf0.0642}&{\bf0.0489}&{\bf0.0407}&{\bf0.0354}&{\bf0.028}&{\bf0.0869}&{\bf0.1291}&{\bf0.1591}&{\bf0.1829}&{\bf0.0653}&{\bf0.1209}\\

% \midrule
% \hline\hline
\thickhline
\multirow{11}{*}{\rotatebox[origin=c]{90}{Spotify}}&POP&0.0111&0.0099&0.009&0.0084&0.0079&0.0009&0.0041&0.0074&0.0101&0.0124&0.0084&0.0107\\

&CP&0.0722&0.058&0.0508&0.0463&0.0435&0.007&0.0261&0.0442&0.0596&0.0741&0.0478&0.0639 \\

&Apriori&0.0189&0.0153&0.0126&0.0114&0.0106&0.0016&0.0064&0.0104&0.014&0.0173&0.0116&0.0155 \\

&Item-item CF&0.0789&0.0672&0.0599&0.0551&0.0515&0.0085&0.0332&0.0564&0.0758&0.0928&0.0579&0.0775 \\

&Prod2Vec&0.1161&0.0896&0.076&0.0677&0.0621&0.0137&0.0485&0.0787&0.1026&0.1232&0.0731&0.1032\\

&\newedit{GraphSAGE}&0.0554&0.0452&0.0396&0.0361&0.0335&0.0052&0.0205&0.0351&0.0474&0.0579&0.0371&0.0497\\

&\newedit{VAE-CF}&0.0669&0.0551&0.0472&0.0422&0.0385&0.0062&0.0248&0.0414&0.0548&0.0662&0.043&0.058\\

&SASRec&0.0307&0.0275&0.0256&0.0243&0.0233&0.0028&0.0119&0.0218&0.0306&0.0389&0.0246&0.0322\\

% &\newedit{GPT4Rec}&0.094&0.084&0.0762&0.0705&0.0661&0.0097&0.0406&0.0705&0.0959&0.1179&0.0733&0.0976\\
&\newedit{BERT4Rec}&0.0083&0.0089&0.0082&0.0078&0.0076&0.0009&0.0041&0.0074&0.0103&0.0134&0.008&0.0106\\
% &\newedit{\textcolor{red}{ComiRec}}&\textcolor{red}{0.0077}&\textcolor{red}{0.0321}&\textcolor{red}{0.0425}&\textcolor{red}{0.0454}&\textcolor{red}{0.0458}&\textcolor{red}{0.0009}&\textcolor{red}{0.0172}&\textcolor{red}{0438}&\textcolor{red}{0.0677}&\textcolor{red}{0.0887}&\textcolor{red}{0.0435}&\textcolor{red}{0.0573} \\
\cmidrule{2-14}
&NPA-SC &{\bf0.1432}&{\bf0.1133}&{\bf0.0938}&\underline{0.0822}&\underline{0.0742}&{\bf0.0164}&\underline{0.0568}&\underline{0.0895}&\underline{0.1145}&\underline{0.1358}&\underline{0.0895}&\underline{0.1207}\\
% &NPA-SC-R &0.1385&0.11&0.0913&0.08&0.0724&0.0161&0.0553&0.087&0.1115&0.1332&0.0874&0.1175\\

&NPA-MC &\underline{0.1419}&\underline{0.1126}&\underline{0.0935}&{\bf0.0823}&{\bf0.0745}&\underline{0.0163}&{\bf0.0573}&{\bf0.0898}&{\bf0.1153}&{\bf0.1368}&{\bf0.0898}&{\bf0.1211}\\

\thickhline
\multirow{11}{*}{\rotatebox[origin=c]{90}{\newedit{Private Industry Dataset}}}&POP&0.1378&0.0549&0.0401&0.0344&0.031&0.0199&0.039&0.0564&0.0719&0.0864&0.0455&0.0774 \\

&CP&0.1378&0.06&0.0438&0.0375&0.0336&0.0199&0.0428&0.0623&0.0793&0.0944&0.0489&0.0825 \\

&Apriori&0.0718&0.0495&0.0433&0.0387&0.035&0.0104&0.0351&0.0609&0.0814&0.098&0.0442&0.0729 \\

&Item-item CF&0.1022&0.0599&0.0456&0.0386&0.0344&0.0156&0.0446&0.0672&0.0847&0.1&0.0498&0.0811 \\

&Prod2Vec&0.0281&0.0179&0.0127&0.0102&0.0087&0.0042&0.0136&0.0192&0.023&0.026&0.0138&0.0219 \\

&\newedit{GraphSAGE}&0.0275&0.0186&0.0142&0.0119&0.0104&0.0043&0.0141&0.0214&0.0268&0.031&0.0153&0.0246\\

&\newedit{VAE-CF}&0.1376&0.0662&0.0491&0.0413&0.0364&0.0201&0.0478&0.0704&0.0882&0.1034&0.0542&0.0893 \\

&SASRec&0.1383&0.0545&0.0394&0.034&0.0309&0.02&0.0387&0.0552&0.071&0.0862&0.0451&0.077\\

% &\newedit{GPT4Rec}&{\bf0.1459}&0.0759&0.057&0.0479&0.0423&\underline{0.0215}&0.0552&0.0825&0.1034&0.1213&0.0621&0.102 \\
&\newedit{BERT4Rec}&0.1362&0.0546&0.0392&0.0329&0.0301&0.0198&0.0384&0.0553&0.0691&0.0838&0.0442&0.0753 \\
% &\newedit{\textcolor{red}{ComiRec}}&\textcolor{red}{0.0499}&\textcolor{red}{0.0438}&\textcolor{red}{0.0372}&\textcolor{red}{0.0331}&\textcolor{red}{0.0302}&\textcolor{red}{0.0073}&\textcolor{red}{0.0322}&\textcolor{red}{0.054}&\textcolor{red}{0.0718}&\textcolor{red}{0.087}&\textcolor{red}{0.038}&\textcolor{red}{0.0626} \\
\cmidrule{2-14}
&NPA-SC &\underline{0.1447}&\underline{0.076}&\underline{0.0572}&\underline{0.0481}&\underline{0.0425}&\underline{0.0214}&\underline{ 0.0558}&\underline{0.083}&\underline{0.1043}&\underline{0.1222}&\underline{0.0623}&\underline{0.1024} \\

&NPA-MC &{\bf0.145}&{\bf0.0776}&{\bf0.0588}&{\bf0.0495}&{\bf0.0436}&{\bf0.0217}&{\bf0.0573}&{\bf0.0858}&{\bf0.1079}&{\bf0.126}&{\bf0.0641}&{\bf0.1047} \\
\bottomrule
\end{tabular}}
\label{table:recommendation_performance}
% \vspace{-2mm}
\end{table*}

\begin{figure*}[bth!]
\centering
\begin{subfigure}{.33\textwidth}
  \centering
\includegraphics[width=1\linewidth ]{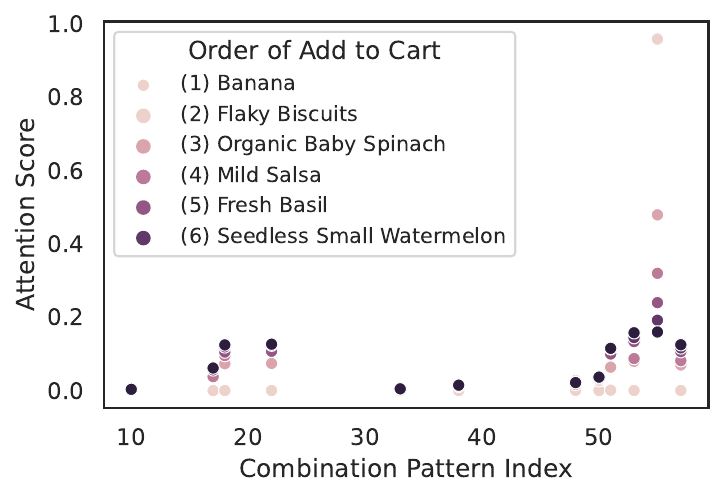}  
\caption{VQA Channel 1}
\label{subfig:multi_intention_head1}
\end{subfigure}
\begin{subfigure}{.33\textwidth}
  \centering
\includegraphics[width=1\linewidth ]{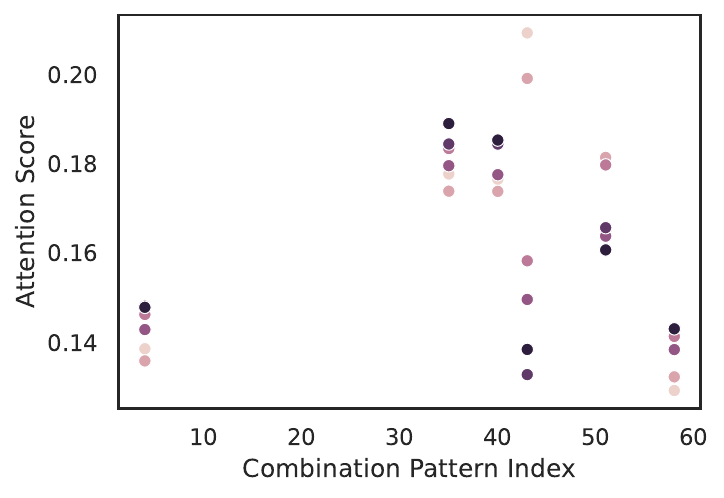}  
\caption{VQA Channel 2}
\label{subfig:multi_intention_head10}
\end{subfigure}
\begin{subfigure}{.33\textwidth}
  \centering
\includegraphics[width=1\linewidth ]{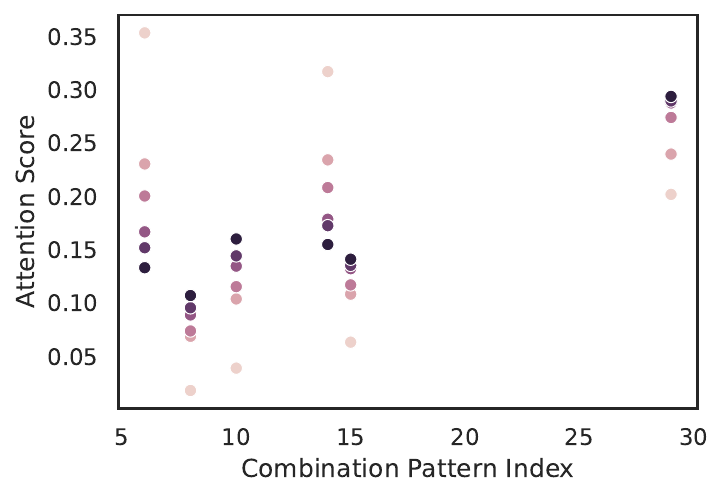}  
\caption{VQA Channel 3}
\label{subfig:multi_intention_head13}
\end{subfigure}
\vspace{-2mm}
\caption{Visualization of the activated combination patterns by different last-layer VQA channels in the NPA framework. Each channel maintains 64 potential combination patterns. Points on the figures show the attention score of a combination pattern when a new item is added to the basket. Item add-to-cart orders are colored and indicated alongside item names in the legend.
% \tina{the legend title is a little bit weird, the item legend could be 1: banana, 2: flaky biscuits, etc. If we mention example of the top scatter point and its dimension, we should label that in the figure.}
}
\label{fig:multi_intention_from_multi_channel}
\vspace{-3mm}
\end{figure*}

\subsection{RQ1: Recommendation Performance}
In this experiment, we conduct a thorough quantitative comparison of 
% within-basket recommendation 
performance between the proposed NPA model and the various baseline models on the three evaluation datasets. 
Here, we include two variations of the proposed NPA model, namely NPA with squashed-context~(NPA-SC) and NPA with multi-context~(NPA-MC) (see Section~\ref{sec:multi_channel_cp_extraction}). 
%\textcolor{red}{as an ablation study}. 
% \textcolor{red}{(1)\textbf{NPA-SC}: The NPA model with squashed-context (SC) mechanism that concatenates sub-contexts estimated in the last layer for producing recommendations. and (2) \textbf{NPA-MC}: The extended NPA model with a multi-context (MC) extraction mechanism as explained in Section~\ref{sec:MMM_NPA}.
% % ~\ref{sec:multi_channel_cp_extraction}.
% }
% \begin{itemize}
%     \item \textbf{NPA-SC}: The NPA model with squashed-context (SC) mechanism. 
%     % \kai{I don't think we talk about SC and MC above. Perhaps we should only list one version of NPA in the paper.}
%     \item \textbf{NPA-MC}: The extended NPA model with a multi-context (MC) extraction mechanism.
%     %,  which treats the identified combination patterns as independent individuals, resulting in multiple contexts that can be used for diversified item recommendations.  
% \end{itemize}

\newedit{Table~\ref{table:recommendation_performance} shows the WBR performance of all models on the three datasets. We conclude that the proposed NPA models perform the best among all methods across all datasets, with 5\%-25\% improvement over the second-best candidates. In particular, NPA-MC outperforms NPA-SC on most datasets and evaluation metrics.}

% One likely reason is that our model does not pose strict assumption that items in baskets are in order and  
% CP is a simple yet strong baseline across all datasets.
% Prod2Vec really good in grocery but bad in spotify.
The sequential models' (i.e., SASRec and BERT4Rec) performance on the Spotify dataset (contains temporal information) is relatively suboptimal, suggesting that over-emphasizing on sequentiality may hurt the generalization of a model on datasets with noisy item orders. This result reflects our motivation that relaxing order sensitivity is beneficial in modeling user behaviors. The NPA models, instead, explicitly model the interleaving intentions, leading to better performance with weak order sensitivity.
% The NPA models' capability of addressing the multiple interleaving intentions issues with the support from combination pattern identification and context extraction while avoiding strong position assumptions
% \tina{users' intention interleaving, with additional support from , -> not over-emphasizing on sequentiality. Order change.} 
% \kai{I think this part needs some work. In this version, NPA uses positional encoding in default which is the same as SASRec and BERT4Rec.} 
GraphSAGE, on the other hand, shows poor performance on all of the datasets. Considering the sparse and noisy nature of those industry datasets, the observation reflects graph models' reliance on extensive interaction data. 
% understandable 
% as it 
% We, therefore, highlight the benefit of modeling combination patterns (as quantized representations in NPA), leading to significantly better performance than other models.} 
It is worth noting that while multiple GNN-based solutions (e.g., BasConv~\cite{liu2020basconv} and MITGNN~\cite{liu2020basket}) were proposed to solve the WBR task, they are incomparable due to their laboratory experiment settings, as also pointed out in~\cite{ariannezhad2023personalized}. 

\begin{figure*}[t!]
\centering
\begin{subfigure}{.32\textwidth}
  \centering
\includegraphics[width=1\linewidth ]{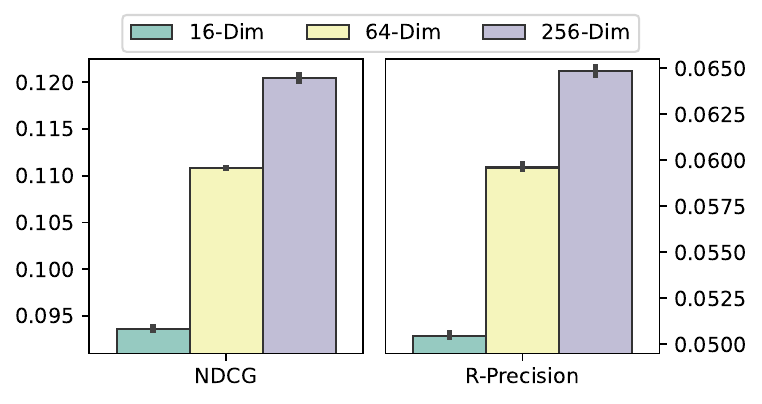}
\caption{Embedding dimensions}
\label{subfig:sensitivity_embedding_dimension}
\end{subfigure}
\begin{subfigure}{.32\textwidth}
  \centering
\includegraphics[width=1\linewidth ]{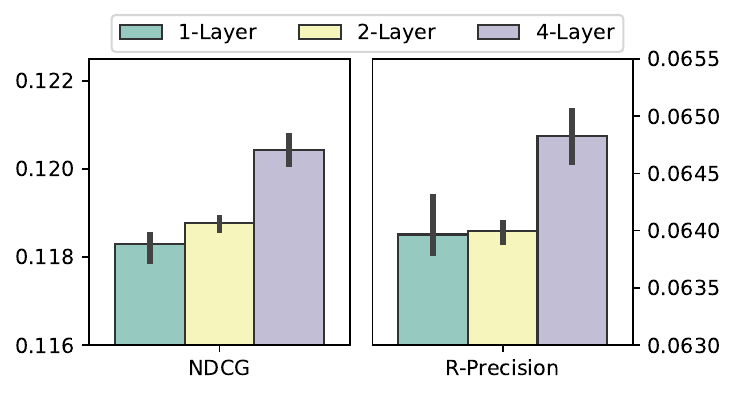}  
\caption{NPA layers}
\label{subfig:sensitivity_block_layers}
\end{subfigure}
\begin{subfigure}{.32\textwidth}
  \centering
\includegraphics[width=1\linewidth ]{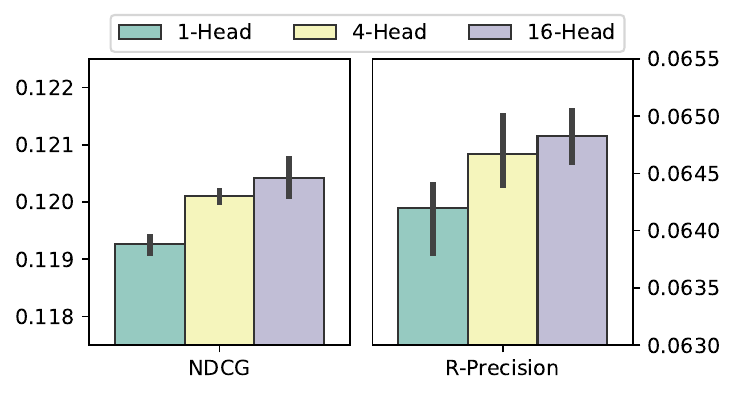}
\caption{Attention heads}
\label{subfig:sensitivity_attention_heads}
\end{subfigure}
\caption{Sensitivity Analysis for NPA-SC on the WBR task on the Instacart dataset. Error bars show 95\% confidence interval.
% \tina{Maybe later fonts to 28} \wuga{This is great plot}
% \tina{highlight multi-granularity in the experiment. Granularity is related with layer, multi-head is related with multi-channel. }
}
\label{fig:sensitivity_analysis}
% \vspace{-3mm}
\end{figure*}

\begin{figure*}[t!]
\centering{
\includegraphics[width=1\linewidth ]{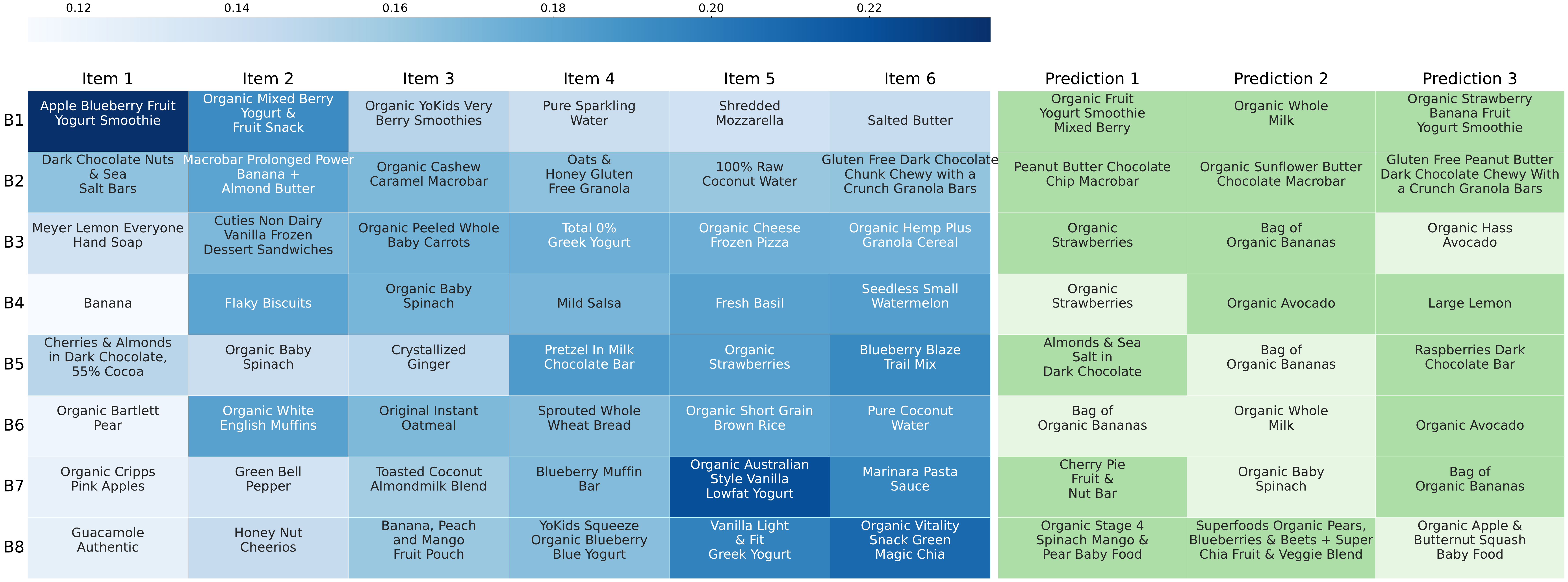}} %\hspace{2mm}
\caption{Attention visualization for Instacart input basket items and 
% the corresponding top predictions generated by the NPA model. 
NPA model's top predictions.
The matrix on the left side illustrates the attention scores for each input item in each basket (row); the darker cell reflects a larger attention weight. The matrix on the right side lists the top recommendation items. The correctly predicted products are highlighted with darker green.
% \kai{Change Input Block X to Input Item X or Input Product X?}
% \wuga{Basket 1-> B1, Input Bock 1 -> Item 1. Don't use warning color for predictions, change it to while-grey. Otherwise, good job.}
}
\label{fig:rec_self_attention}
% \vspace{-4mm}
\end{figure*}

% \begin{figure*}[t!]
% \centering{
% \includegraphics[width=1\linewidth ]{figs/RQ3/RQ3_fig_June30.pdf}} %\hspace{2mm}
% \caption{Attention visualization for Instacart input basket items and the corresponding top predictions generated by the NPA model. The matrix on the left side illustrates the attention scores for each input item in each basket (row); the darker cell reflects a larger attention weight. The matrix on the right side lists the top recommendation items \tina{maybe highlight hits}.
% % \kai{Change Input Block X to Input Item X or Input Product X?}
% % \wuga{Basket 1-> B1, Input Bock 1 -> Item 1. Don't use warning color for predictions, change it to while-grey. Otherwise, good job.}
% }
% \label{fig:rec_self_attention_june30}
% %\vspace{-3mm}
% \end{figure*}

\subsection{RQ2: Capturing Shift of Shopping Intentions by the NPA Framework}
\label{RQ2_capture_shopping_intention_shifts}
In this section, we analyze how the multiple intentions of shopping baskets are captured by an NPA model.  Specifically, given the items added to a shopping cart (using Instacart data), we show the distribution shift of the activation of combination patterns on different last-layer VQA channels and present the result in Figure~\ref{fig:multi_intention_from_multi_channel}.

To begin, the non-uniform distribution of attention scores across combination pattern dimensions indicates the presence of multiple intentions within shopping baskets, with varying degrees of activation across these patterns.
% Secondly, the distribution of the attention scores across the combination pattern dimensions are different between the sampled self-attention heads. Reflecting the model's capability of estimating the overall context of a shopping basket based on multiple aspects captured by the multiple context channels.
In addition, the differing distribution of attention scores among VQA channels reflects the model's ability to assess a shopping basket's overall context by integrating facets derived from multiple context channels. 

Interestingly, with new items added to the basket, the attention distribution on combination pattern changes, which reflects the user's intention changes during shopping. In particular, attention changes are not necessarily monotonous; such a phenomenon shows the back and forth of user intention (interleaving).

% In addition, combination pattern dimension could be interpreted by analysing their activation magnitude attributed by input items, we provide such analyses in Appendix~\ref{app:wordcloud}.} 
% For example, the top right point in Figure~\ref{subfig:multi_intention_head1} indicates "Banana`` has a near 100\% activation of the corresponding combination pattern while adding latter items in the basket decreases the attention score on the specific dimension.
% \tina{We provide additional interpretation of combination patterns in Appendix.} 

\subsection{RQ3: Sensitivity Analysis and Ablation Study}
In this section, we study the sensitivity of NPA's performance under different hyperparameter settings, and evaluate the NPA model performance with either the multi-layer or the multi-channel settings removed (i.e., a one-layer or one-head model variation) as an ablation study. This analysis includes three hyperparameter settings, 1) the size of the model's embedding dimensions, 2) the number of NPA layers for higher-level context extraction described in Section~\ref{sec:MMM_NPA},
% ~\ref{sec:shared_multi_context},
and 3) the number of channels (attention heads) within the VQA modules for sub-context extraction described in Section~\ref{sec:MMM_NPA}.
% ~\ref{sec:multi_channel_cp_extraction}.
% and 4. the per-training step model performance. \kai{Let's drop 4.} 
Figure~\ref{fig:sensitivity_analysis} shows the analysis results. 
% \textcolor{red}{along with 95\% confidence interval.}
% , where the error bar denotes a 95\% confidence interval. 
%First of all, 

Figure~\ref{subfig:sensitivity_embedding_dimension} shows a strong positive correlation between the embedding dimensions and model performance. This suggests that the Instacart dataset requires 
% the NPA model to have 
more encoding allowance for the latent information of combination patterns for effective recommendations,
while such an effect might saturate at a certain level. 
% Similar observations can also be obtained on other datasets\tina{suggested to delete the prev sentence; so if we are not showing results in the unlimited-spacing appendix section, then we should probably not mention this.}. 
%In addition, results from 

Figure~\ref{subfig:sensitivity_block_layers} suggests that the model performance correlates positively with the number of NPA layers; having more NPA layers helps to extract the combination pattern information from the multiple granularity levels, reflecting the necessity of considering the larger magnitude of multi-granularity into the model. 
The overlapping 95\% confidence interval on Figure~\ref{subfig:sensitivity_attention_heads} shows that although having multiple attention heads increases model performance compared to when only a single attention head is used for combination pattern identification, the effect of increased attention heads saturates at a certain level (i.e., 16 heads for this example). This observation suggests an intuitive understanding of within-basket recommendation tasks; while multiple combination patterns could exist in the shopping baskets, there should be an appropriate upper limit for possible combination patterns for different datasets (e.g., user's overall intentions during a single grocery shopping session are not expected to be more than 10). 

Results from Figure~\ref{subfig:sensitivity_block_layers} and~\ref{subfig:sensitivity_attention_heads} reflect the necessity of utilizing the multi-layer, multi-channel NPA model to address the problem of multi-granular shopping intentions.

\subsection{RQ4: Recommendation Explanation via In-basket Item Attentions}
\label{sec:recommendation_explanation_via_item_attention}
% As mentioned in Section~\ref{sec:prediction_explanation_context_attention}, the NPA model is self-interpretable by merging the context attentions of each layer. 
%  Specifically, for the recommended products given the existing items in a basket, the underlying attention onto the input basket for producing such recommendations can be extracted and visualized for explaining the recommendations. 
In this section, we perform experiments on the Instacart dataset for recommendation interpretations 
% through attention visualization of the corresponding input basket items.
via attention visualization of input basket items.
The experiment results are illustrated in Figure~\ref{fig:rec_self_attention}.
% , where the input baskets are represented as rows in the matrix on the left, with attention scores assigned to each item. The matrix on the right displays the top recommendations for each basket.

\newedit{Based on the results, the attention scores assigned to basket items do not follow a strictly increasing or decreasing pattern.}
% Based on the results presented, it can be observed that the distribution of attention scores assigned to the basket items does not exhibit a strictly increasing or decreasing pattern.
% First of all, the observations of the illustrated results show that the distribution of attentions onto the basket items do not strictly follow a sequentially increasing or decreasing pattern. 
\newedit{For instance, in B1, the attention scores peak at item 1, while the distribution in B2 appears to be more uniform.} 
% For example, the attention scores assigned to the items in basket B1 exhibit a peak at item 1, whereas the distribution for B2 appears to be more uniform. 
% the attention score distribution for B1 shows a peak at item 1, while that for B2 is more evenly distributed. 
\newedit{Conversely, the model assigns lower attention scores to the first item in baskets B3 (i.e., \emph{Hand Soap}) and B4 (i.e., \emph{Banana}), considering them as noisy inputs unrelated to the rest of the basket.}
% In contrast, the model assigns lower attention scores to the first item in baskets B3 and B4, as these items are considered noisy inputs that are unrelated to the rest of the basket. 
% In particular, \emph{Hand Soap} and \emph{Banana} are deemed to have little relevance for context estimation.
% At the same time, the model puts less attention onto the first item for B3 and B4, where the ignored ``Hand Soap" and ``Banana" would be considered as noisy inputs for context estimation and in fact they are relatively unrelated to the rest of the basket. 
% \tina{we should mentioned noisy items somewhere before in the paper.}
In B8, the attention scores highlight two primary themes: healthy snacks for adults (e.g., \emph{yogurt}, \emph{chia}, \emph{fruit pouch}) and 
baby foods (i.e., \emph{YoKids squeeze yogurt}). 
Consequently, the model generates recommendations aligning with these themes, such as baby food and fruit \& veggie blends.
% , which help to complete the basket.
% \kai{I am not sure if these products can be categorized into ``drinks''. I would say they are ``healthy snacks''}
% for both babies and adults. 
% \lan{The YoKids squeeze organic blueberry yogurt might not make a smoothie - The yogurt comes in a convenient pouch with a twist-off cap that makes it easy for children to eat on the go. This item might indicate the user has baby not fair to say the baby also drinks the healthy drink. It has been suggested baby could have this kind of yogurt after 6 months, which aligns with our prediction - stage 4 baby food}
% The corresponding recommendations suggests that the model has successfully identified both combination pattern by recommending 
% fruity baby food and healthy organic blend.

\newedit{Furthermore, 
an evenly distributed set of attention scores does not necessarily indicate the existence of multiple combination patterns.}
% a more evenly distributed set of attention scores does not necessarily imply the presence of multiple combination patterns. 
\newedit{For instance, 
% basket B1 exhibits a peak attention score for yogurt smoothies, whereas the attention scores for B2 are more evenly distributed. 
% \newedit{despite the difference of attention score distributions in B1 and B2}, 
% the model generates smoothie-related recommendations for B1, whereas the recommendations for B2 suggest that the model has identified a general intent of finding chocolate macrobars or granola bars.
despite the varying attention score distributions observed across baskets B1 and B2, the model is able to identify one general intention behind each basket and provide relevant recommendations (smoothie-related products for B1, chocolate macrobars/granola bars for B2).} 
These findings suggest that the NPA model is capable of generating satisfactory and explainable recommendations using its cascaded context attention.

\section{Conclusion}
In this paper, we presented Neural Pattern Associator (NPA) to address the within-basket recommendation challenges. The core computation unit of NPA is the novel Vector Quantized Attention module, which can explicitly model combination patterns (certain shopping combinations) and contexts of incomplete baskets. With multi-layer, multi-channel, and multi-step context extractions, the NPA model can further capture higher-level context information, allowing it to recommend items to a shopping basket while preserving reasonable coherency. Through extensive experiments, we demonstrated the effectiveness of NPA compared to multiple well-known baselines. 
%in terms of various performance metrics. 
%We also visualized the attention distributions of the NPA model on the incomplete baskets along with its top recommendations,
Through extensive ablation study and visualization, we show that the model can capture the complex relationships among items in a basket and provide interpretable recommendations. 
\bibliographystyle{ACM-Reference-Format}
\bibliography{WWW22}

% \newpage
\appendix
\begin{center}\LARGE \textbf{Appendix}\end{center}

\section{Neural Pattern Associator with the Multi-context (MC) Mechanism }
\label{app:npa_mc_details}
\begin{figure}[t]
    \centering
        \includegraphics[width=0.95\linewidth]{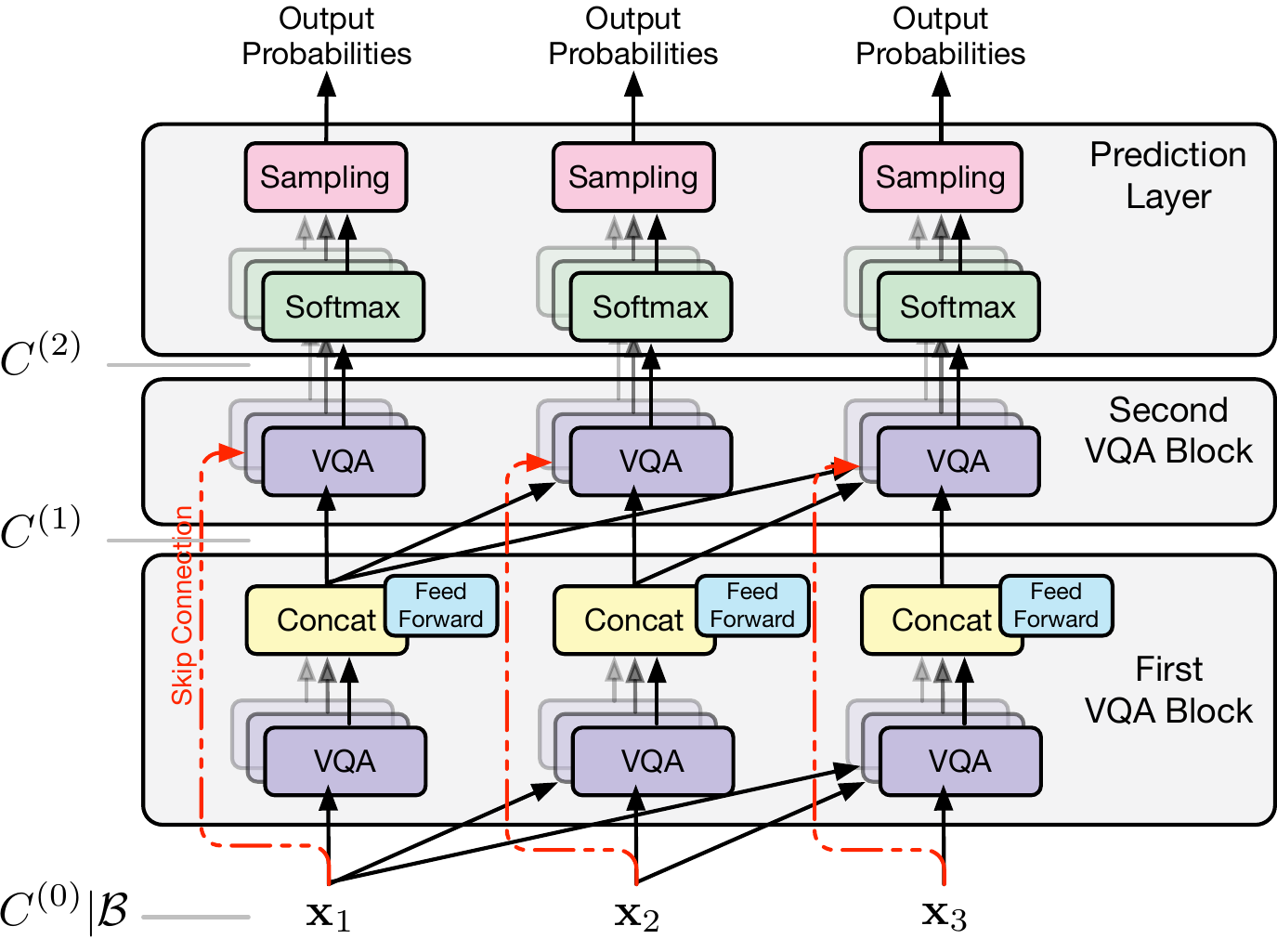}
    \caption{The high-level overview of Neural Pattern Associator model (NPA-MC). The figure shows a two-layer NPA with three channels, where multiple VQA computation units are stacked in a similar fashion of multi-head attention. }
\label{fig:npa-mc}
\vspace{-2mm}
\end{figure}

As mentioned in Section~\ref{sec:multi_channel_cp_extraction}, NPA-MC is an extended version of the NPA framework where it uses the multi-context extraction mechanism.
We additionally illustrate the model architecture of a two-layer, three-channel NPA-MC in Figure~\ref{fig:npa-mc}. Different from the architecture of a similar-structured NPA-SC model demonstrated in Figure~\ref{fig:npa}, the last layer of NPA-MC treats each $\mathbf{c}^h$ as independent context estimation and performs Gumbel sampling~\cite{jang2016categorical, maddison2016concrete} to extract different combination patterns that result in different contexts.
%\tina{let's try to avoid using ``next product" as we claimed that we're trying to complete a basket with multiple groud truth.}

\subsection{WBR in a Multi-context (MC) Setting}
\label{sec:multi_context_setting}
The direct consequence of producing multiple contexts for a basket is that the product scoring function described in Equation~\eqref{eq:single_score} is no longer valid for relevant product searches. While one can aggregate scores estimated from the individual context in the form 
% \tina{$\mathbf{c}$ below should be $\mathbf{c}^h$ ?}
\begin{equation}
    p_{\vartheta}(\mathbf{x}_j|C) = \frac{1}{|C|}\sum_{\mathbf{c}^h\in C} p_{\vartheta}(\mathbf{x}_j|\mathbf{c}^h),
\end{equation}
we observe a significant popularity bias due to the over-amplification of product prioritization, particularly for products that are marginally relevant to all contexts. For example, bananas in the Instacart dataset are recommended for almost all baskets. We address this issue by leveraging the Free Energy Scoring Function (FESF)~\cite{wang2022watermarking} that merges score estimations of all contexts such that
\begin{equation}
    \log p_{\vartheta}(\mathbf{x}_j|C) = \log\sum_{\mathbf{c}^h\in C}\exp(\mathbf{e}_j^{\top}\mathbf{c}^h) / T,
\end{equation}
where $T$ is the temperature that controls the polarization of the score estimation. In addition, $p_{\vartheta}(\mathbf{x}_j|C)$ is not a valid conditional probability as the score is no longer bounded by a partition function. FESF fits our purpose of prioritizing products that satisfy the ranking conditions described in Section~\ref{sec:prob_state}. 
%\kai{The current implementation doesn't use FESF.}
% lets stick to this, since addtional layer of novelty will introduce more confusion. 

\section{Training Objective of NPA for Data without Temporal Information}
\label{app:training_obj_wo_temporal_info}
% In contrast to Section~\ref{sec:temporal_data}, 
For the dataset without temporal information, we skip the position encoding and sample multiple random permutations of the products in a basket to maximize data utilization, resulting in an Any-Order Autoregressive Model (AO-ARM) such that
% \kai{Is this equation too long? the reference number is below the equation.} \tina{fixing it to have the reference number beside would increase the space it takes than the current approach. }
% \begin{fleqn}[\parindent]
\begin{equation*}
\begin{split}
    \argmax_{\Theta}\!\sum_{o} \!\left[
    \! E_{\mathbf{s}^{(o)}\sim \textit{Perm}(\mathcal{B}^{(o)})}
    \!\left[\!
    %\vphantom{\int_1^2}  % spacing trick to display the large [.
    \sum_{t=2}^{|T_\mathcal{B}^{(o)}|}\!\log p_{\Theta}\left(s^{(o)}_{t}|\mathbf{s}^{(o)}_{<t}\right)\! \right]\!\right]\!.
\end{split}
\end{equation*}
% \end{fleqn}
With the stochastic gradient descent training approach, the trained model will converge to be order-invariant due to observing arbitrarily random orders of products in a basket, enabling efficient inference on all possible conditionals. 

In a similar fashion to Equation~\ref{eq:obj_no_temporal_info}, for the multi-context NPA model, we extend random ordered training objective as follows
% \kai{Is this equation too long? the reference number is below the equation.}\tina{fixing it to have the reference number beside would increase the space it takes than the current approach. }

\begin{equation}
\begin{split}
    & 
    \argmax_{\Theta} \sum_{o} \left[ 
    E_{\mathbf{s}^{(o)}\sim \textit{Perm}(\mathcal{B}^{(o)})} 
    \left[ \vphantom{\int_1^2}  % spacing trick to display the large [.
    \right. \right. 
    \\
    & 
    \left. \left. \sum_{t=2}^{|T_\mathcal{B}^{(o)}|} \max_{\mathbf{c}_t^{(o)}\in C_t^{(o)}} 
        \log p_{\vartheta}(s^{(o)}_{t}|\mathbf{c}_t^{(o)}) p_{\theta}(\mathbf{c}_t^{(o)}|\mathbf{s}^{(o)}_{<t})
        \right]
        \right],  
\end{split}
\end{equation}

% \begin{equation}
% % \resizebox{0.45\textwidth}{!}{%$
% \argmax_{\Theta}\sum_{o} \left[E_{\mathbf{s}^{(o)}\sim \textit{Perm}(\mathcal{B}^{(o)})}[\sum_{t=2}^{|T_\mathcal{B}^{(o)}|} \max_{\mathbf{c}_t^{(o)}\in C_t^{(o)}} 
%         f( s^{(o)}_{t}, \quad 
%         \mathbf{s}^{(o)}_{<t},  \quad 
%         \mathbf{c}_t^{(o)})
%     ]\right],
%     % $%}
% \end{equation}
% where $f\left( s^{(o)}_{t}, \mathbf{s}^{(o)}_{<t}, \mathbf{c}_t^{(o)} \right) = \log p_{\vartheta}\left(s^{(o)}_{t}|\mathbf{c}_t^{(o)}\right) p_{\theta}\left(\mathbf{c}_t^{(o)}|\mathbf{s}^{(o)}_{<t}\right)$

While the training objective looks complex with first glance, it is in the same format of training a Transformer model with an additional operation that conducts max-pooling before merging losses from each time step.

\section{Experiment Setting Details}
\label{app:experiment_settings}
\subsection{Dataset Descriptions}
\label{app:dataset_descriptions}
While Table~\ref{tb:datasets} lists out the summary of basic statistics of the datasets we use in the experiments, we additionally provide detailed descriptions of these datasets below:
\begin{itemize}
    \item \textbf{Instacart.}
    Instacart offers same-day grocery delivery in the US and Canada. The dataset comprises approximately 50,000 grocery products and 3,000,000 orders. Each product includes contextual information such as name, aisle, and department. User contexts are unavailable, but the order of user activities is recorded.
    % Instacart provides an online service for same-day grocery delivery in the US and Canada. The dataset contains around 50,000 grocery products 
    % with shopping records of around 200 thousand users, 
    % with a total of around 3,000,000 orders. All products have contextual information, including name, aisle, and department. User contexts are not available. Time information is not presented, but the order of user activities is available. 
    %We remove carts with less than 7 items. For model training and inference, we use 6 items as input.
    \item \textbf{Spotify.} 
    The Spotify Million Playlist dataset consists of 1,000,000 playlists created by users on the Spotify platform between January 2010 and October 2017. Each playlist contains the playlist title, the track list (including track metadata), and other miscellaneous information 
about the playlist. The position of the track in the playlist is also included.
    % The Spotify Million Playlist dataset contains 1,000,000 playlists, including playlist titles and track titles, created by users on the Spotify platform between January 2010 and October 2017. 
    % It consists of 817,741 transactions. We treat transactions that happened on the same day and are made by the same user id as one cart. After grouping by transaction date and user id, there are 119,578 carts.
    %We remove playlists with less than 11 tracks and tracks occurring in less than one hundred playlists in total. For model training and inference, we use 10 items as input.
    % \item \textbf{Tafeng.} Tafeng contains a Chinese grocery store's transaction data from November 2000 to February 2001 
    % released by ACM RecSys \kai{I can't confirm if it was indeed released by RecSys. So I comment this out.} 
    %We remove carts with less than 7 items. For model training and inference,  we use 6 items as input.
    % the records of over 13000 before filtering.
    
    % \kai{This number is probably before filtering. Not sure what 13000 refers to. Need to double check with Tina.}  
    \item \textbf{Private Industry Dataset.\footnote{Dataset information is suppressed for triple-blind review.}} A private industry dataset comes from a large North American food retailer. All baskets contain at least 11 items and no more than 40 items. The add-to-cart order is not present.
    % \kai{"Industry Dataset"} 
    %For model training and inference, we use 10 items as input.
\end{itemize}

\begin{table*}[t]
\centering
\caption{Summary of datasets. 
% \tina{The users statistics have been moved to Section ~\ref{sec:appdx_personalization}, delete the "\#Users" column in this table.} 
% \wuga{ consider split the avg+std colum into two columns with respective values. Compute sparsity as additional column, which is Carts-Items matrix sparsity.}
}
% \kai{sparsity can be seen as average cart length divided by number of items}
% \rowcolors{2}{gray!10}{white}
\resizebox{0.95\linewidth}{!}{%
\begin{tabular}{c|ccccccc}
\toprule
Dataset & \newedit{Application} & \# Items & \# Baskets & \makecell{Avg. Length / Basket} & \makecell{Std. Length / Basket} & Sparsity\\
\midrule
Instacart & \newedit{Grocery Basket
Completion} & 49,276 & 1,148,912 & 14.0822 & 7.0387 & $2.86 \times 10^{-4}$\\
Spotify & \newedit{Music Playlist Extension} & 69,887 & 882,207 & 59.8522 & 46.2714 & $8.56 \times 10^{-4}$\\
% Tafeng & 22,730 & 44,388 & 13.2562 & 7.4242 & $5.83 \times 10^{-4}$\\
\newedit{Private Industry Dataset} & \newedit{Grocery Basket
Completion} & 55,060 & 570,632 & 24.8266 & 8.0093 & $4.51 \times 10^{-4}$\\
\hline
\end{tabular}%
}
\label{tb:datasets}
% \vspace{-5mm}
\end{table*} 

\begin{figure*}[t!]
\centering
\begin{subfigure}{.33\textwidth}
  \centering
\includegraphics[width=1\linewidth ]{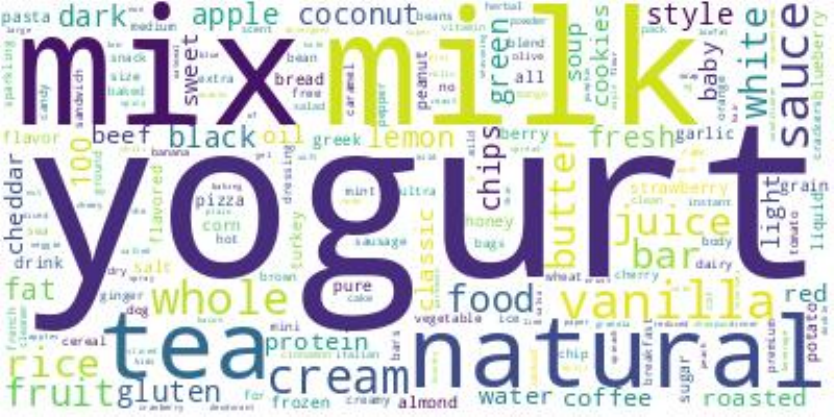}
\caption{Codebook Entry 1}
\label{subfig_cp1}
\end{subfigure}
\begin{subfigure}{.33\textwidth}
  \centering
\includegraphics[width=1\linewidth ]{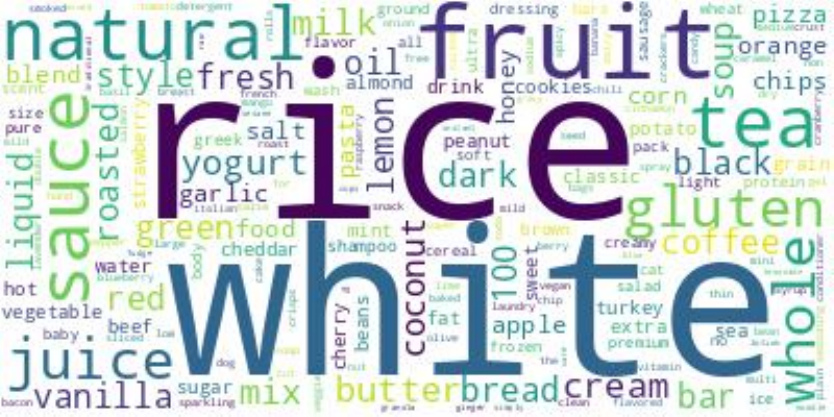}
\caption{Codebook Entry 2}
\label{subfig_cp2}
\end{subfigure}
\begin{subfigure}{.33\textwidth}
  \centering
\includegraphics[width=1\linewidth ]{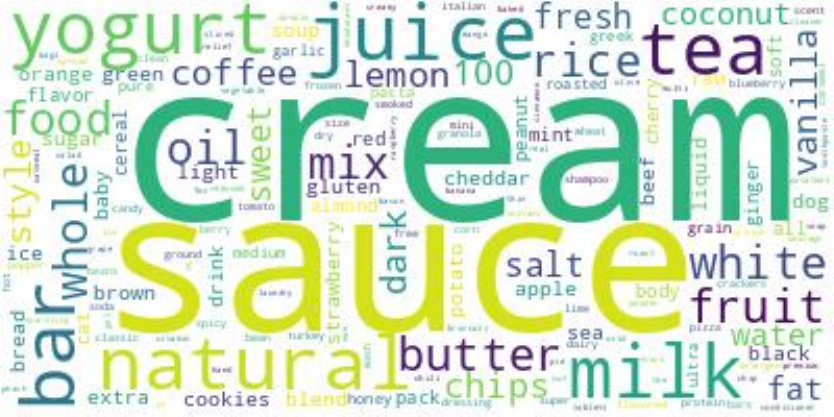}
\caption{Codebook Entry 3}
\label{subfig_cp3}
\end{subfigure}
\caption{Word clouds of top words that activate each of the combination patterns in the shared codebook $Z^L$ for NPA model trained on Instacart dataset. 
}
\label{fig_wordCloud_combination_patter_to_item}
\end{figure*}

\subsection{Model Implementation Details}
\label{app:model_implementations}
We describe the model implementation details in this section.
\newedit{
% For SASRec\footnote{https://github.com/kang205/SASRec} and BERT4Rec\footnote{https://github.com/FeiSun/BERT4Rec}, we use code provided by the corresponding authors. 
To ensure a fair and rigorous comparison, we employ the Mlxtend\footnote{https://rasbt.github.io/mlxtend/} and PyG (PyTorch Geometric)~\cite{Fey/Lenssen/2019} libraries for implementing the Apriori and GraphSAGE algorithms, respectively. Additionally, we utilize PyTorch to implement Item-item CF, Prod2Vec, and VAE-CF. For the SASRec and BERT4Rec models, we utilize the code provided by the respective authors.
Among the related models, several hyperparameters are commonly shared. These include the embedding dimension size, chosen from \{16, 32, 64, 128, 256\}, the $\ell_2$ regularizer selected from \{1, 0.1, 0.01, 0.001, 0.0001, 0.00001\}, and the dropout rate chosen from \{0, 0.1, 0.2, ..., 0.9\}. Other hyperparameters, such as the mask proportion in BERT4Rec and initialization strategies, are either tuned in a similar manner to the NPA model or set according to the instructions provided in the original papers, if available. The results of each baseline model are reported using their respective optimal hyperparameter settings.}

\newedit{For training the NPA model, we adopt the \emph{AdamW} optimizer~\cite{loshchilov2017decoupled} with learned positional embeddings, a learning rate of 0.0003, a batch size of 256, a dropout rate of 0.1, and a codebook size of 64. 
% In the case of 
\newedit{Under the multi-context setting (see Section~\ref{sec:multi_context_setting}),}
% NPA-MC, 
the head number in the last layer is set to 5. We also utilize residual connections to stabilize model training. After tuning, we select 6 layers for the Spotify dataset and 4 layers for the remaining two datasets. The optimal head size for lower layers is determined as 16 for the Instacart dataset and 32 for the other two datasets. The NPA model is implemented using PyTorch. All models are trained from scratch on a single NVIDIA A100 GPU.
% We tune the number of layers and head number using the validation set, resulting in 
}

\section{Inspecting Combination Patterns in the Codebook}
\label{app:wordcloud}

As described in Section~\ref{sec:simple_combination_pattern}, the VQA module is designed for explicitly modeling
the combination patterns of the input basket. 
In this section, we conduct an experiment to inspect the information captured by the combination pattern codebook (last layer shared codebook $Z^L$). 
To do so, we visualize the word frequency of the input item names that activate each combination pattern as word clouds, as shown in Figure~\ref{fig_wordCloud_combination_patter_to_item}.
Figure~\ref{subfig_cp1} illustrates an example combination pattern, which is most likely to be activated by items that are related to healthy drinks such as ``yogurt'', ``milk'', and ``tea''.
The combination pattern example illustrated in Figure~\ref{subfig_cp2} suggests that this combination pattern is more likely to be activated by items related to ``white rice'' and ``fruit'', while the combination pattern presented in Figure\ref{subfig_cp3} is more related to cooking products such as ``sauce'', ``cream'', ``milk'', ``butter'', etc. We note that since each combination pattern in the codebook is not considered to be completely disentangled, items that activate these combination patterns are not completely diversified. %Overall, we show that the combination patterns learned by the NPA model is explainable, and can help to identify the potential multiple combination patterns within an incomplete basket.

\begin{figure*}[t!]
\centering
\begin{subfigure}{.33\textwidth}
  \centering
\includegraphics[width=1\linewidth ]{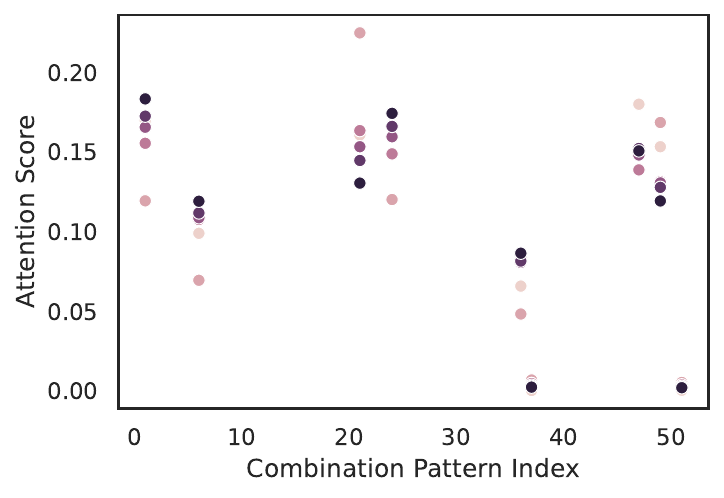}  
\caption{Sampled Basket 1 Channel 1}
% \label{subfig:multi_intention_head1}
\end{subfigure}
\begin{subfigure}{.33\textwidth}
  \centering
\includegraphics[width=1\linewidth ]{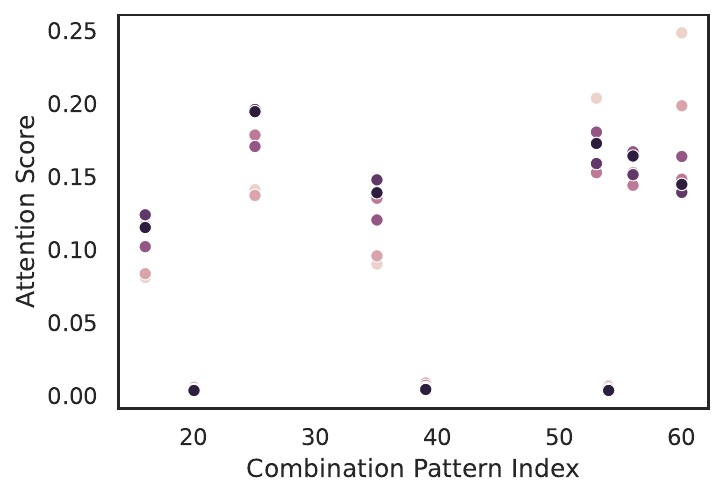}  
\caption{Sampled Basket 1 Channel 2}
% \label{subfig:multi_intention_head10}
\end{subfigure}
\begin{subfigure}{.33\textwidth}
  \centering
\includegraphics[width=1\linewidth ]{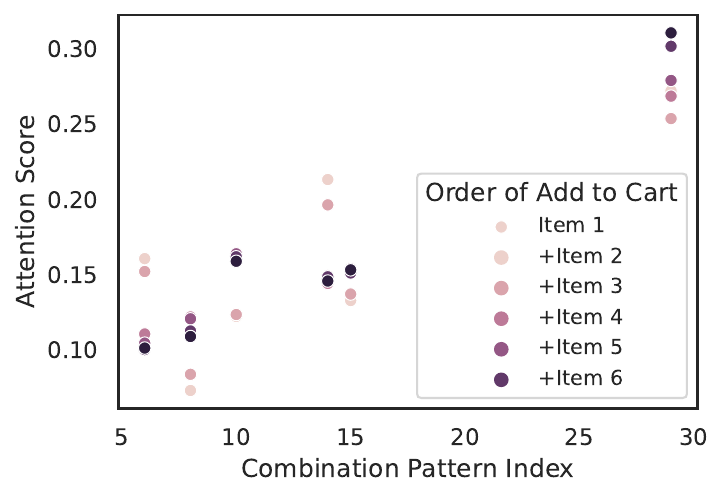}  
\caption{Sampled Basket 1 Channel 3}
% \label{subfig:multi_intention_head13}
\end{subfigure}
% basket 2
\begin{subfigure}{.33\textwidth}
  \centering
\includegraphics[width=1\linewidth ]{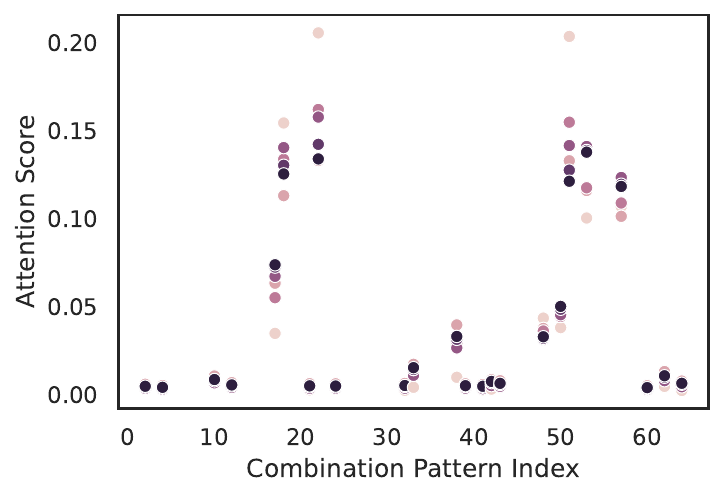}  
\caption{Sampled Basket 2 Channel 1}
% \label{subfig:multi_intention_head1}
\end{subfigure}
\begin{subfigure}{.33\textwidth}
  \centering
\includegraphics[width=1\linewidth ]{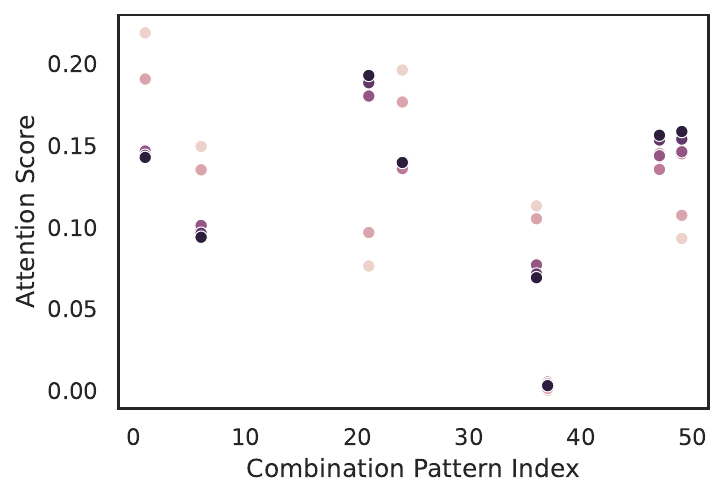}  
\caption{Sampled Basket 2 Channel 2}
% \label{subfig:multi_intention_head10}
\end{subfigure}
\begin{subfigure}{.33\textwidth}
  \centering
\includegraphics[width=1\linewidth ]{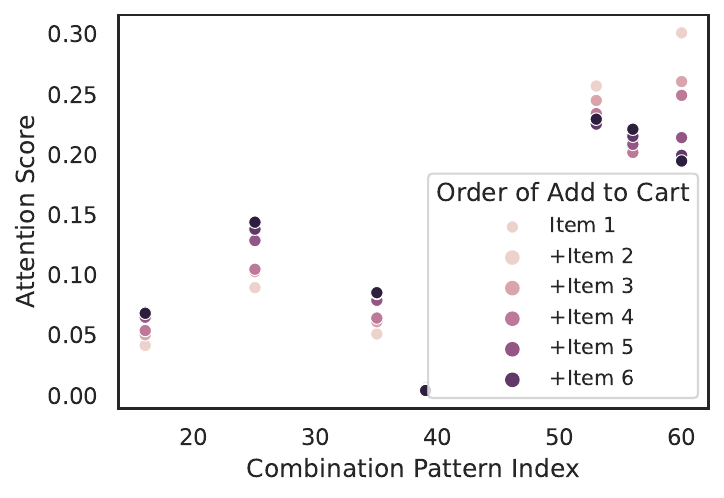}  
\caption{Sampled Basket 2 Channel 3}
% \label{subfig:multi_intention_head13}
\end{subfigure}
% basket 3
\begin{subfigure}{.33\textwidth}
  \centering
\includegraphics[width=1\linewidth ]{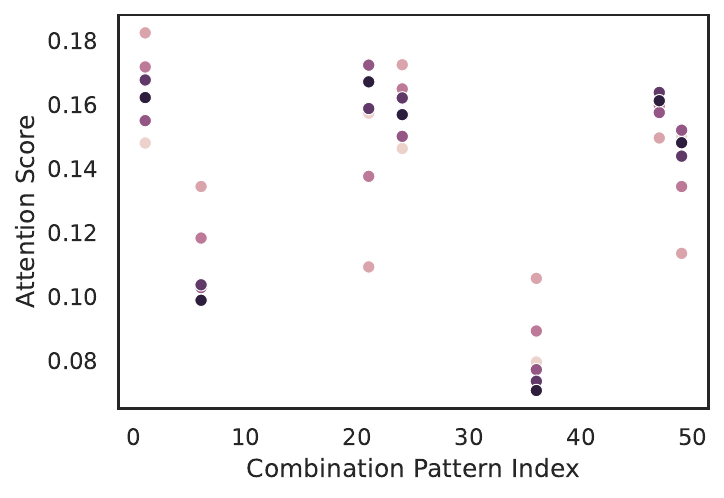}  
\caption{Sampled Basket 3 Channel 1}
% \label{subfig:multi_intention_head10}
\end{subfigure}
% \begin{subfigure}{.33\textwidth}
%   \centering
% \includegraphics[width=1\linewidth ]{figs/RQ_Multi_Intention/B7/attention_head11.pdf}  
% % \caption{HEAD 10}
% % \label{subfig:multi_intention_head13}
% \end{subfigure}
% \\
\begin{subfigure}{.33\textwidth}
  \centering
\includegraphics[width=1\linewidth ]{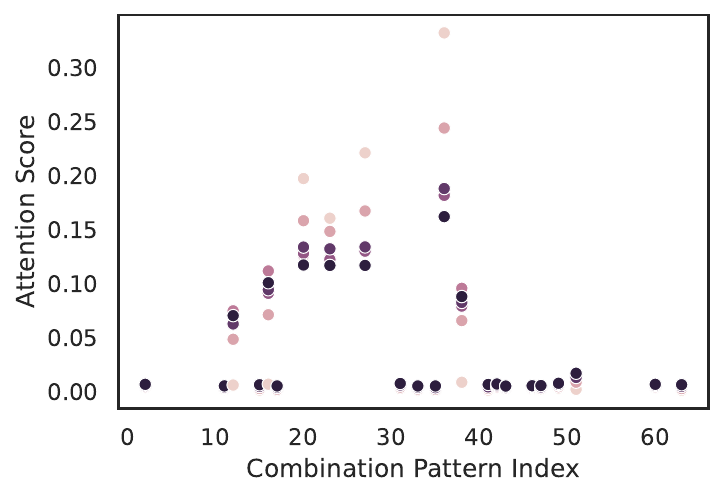}  
\caption{Sampled Basket 3 Channel 2}
% \label{subfig:multi_intention_head13}
\end{subfigure}
\begin{subfigure}{.33\textwidth}
  \centering
\includegraphics[width=1\linewidth ]{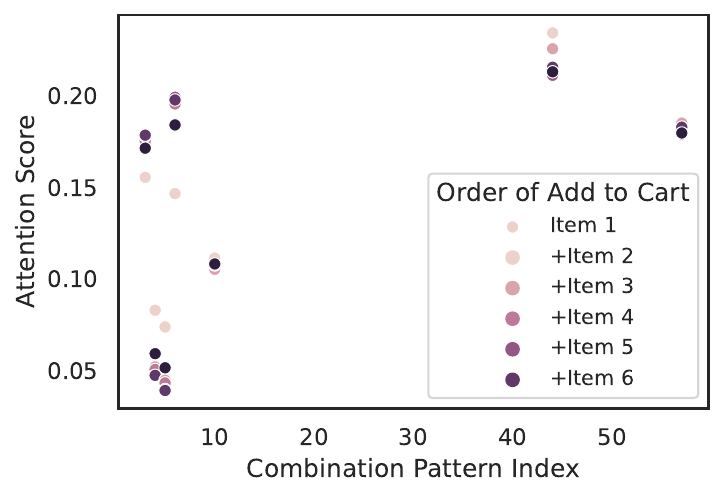}  
\caption{Sampled Basket 3 Channel 3}
% \label{subfig:multi_intention_head13}
\end{subfigure}
\caption{Visualization of the activated combination patterns by different last-layer VQA channels in the NPA framework. Each channel maintains 64 potential combination patterns. Points on the figures show the attention score of a combination pattern when a new item is added to the basket. Item add-to-cart orders are colored and indicated alongside item names in the legend.}
\label{fig:additional_activated_cp_visualization}
\end{figure*}

\section{Additional illustrations of shopping intention shifts captured by the NPA Framework}
As discussed in Section~\ref{RQ2_capture_shopping_intention_shifts}, an NPA model can capture the multiple intentions of shopping baskets, which can be illustrated by showing the distribution shift of activation of combination patterns on different last-layer VQA channels. We additionally provide such visualizations of multiple sampled input (Instacart) baskets and sample channels in Figure~\ref{fig:additional_activated_cp_visualization}.

\end{document}